\documentclass[twocolumn]{aastex631}

\usepackage{amsmath}
\usepackage{graphicx}

\usepackage{natbib}
\usepackage{xcolor}
\usepackage{booktabs}
\usepackage{tabularx}
\usepackage{enumitem}

\usepackage{multirow}
\usepackage{hyperref}

\begin{document}
\title{Assessing the Vera Rubin Observatory’s Ability to Discover Asteroid Impactors Before They Collide with Earth}

\author[0009-0002-7847-8082]{Qifeng Cheng}
    \affiliation{Department of Physics, Duke University, Durham, NC 27708, USA}
\author[0000-0002-4934-5849]{Daniel Scolnic}
      \affiliation{Department of Physics, Duke University, Durham, NC 27708, USA}
      \affiliation{Department of Electrical and Computer Engineering, Duke University, Durham, NC 27708, USA}
\author[0009-0005-5452-0671]{Jacob A. Kurlander}
      \affiliation{DiRAC Institute and the Department of Astronomy, University of Washington, 3910 15th Ave NE, Seattle, WA 98195, USA}
\author[0009-0005-9428-9590]{Ian Chow}
      \affiliation{DiRAC Institute and the Department of Astronomy, University of Washington, 3910 15th Ave NE, Seattle, WA 98195, USA}
\author[0000-0002-9561-9249]{Maryann Benny Fernandes}
    \affiliation{Department of Electrical and Computer Engineering, Duke University, Durham, NC 27708, USA}

\begin{abstract}
Asteroid impactors larger than $\sim$10 m, from Chelyabinsk-scale airburst and Tunguska-scale events to $>$300 m continental threats, remain the dominant planetary-defense risk. While the Vera C.~Rubin Observatory Legacy Survey of Space and Time (LSST) will transform Solar System science, its observing cadence and survey design were not specifically optimized to discover imminent impactors. To assess its performance, we introduce a new method for efficiently generating synthetic impactor populations by minimally perturbing sampled \texttt{NEOMOD3} orbits and evaluate their discovery efficiency with the \texttt{Sorcha} survey simulator. Our simulations show that LSST discovers 79.7\% of large impactors ($>$140 m), decreasing to 50.3\% for upper mid-sized (50-140 m), 26.8\% for lower mid-sized (20 - 50 m), and 10.5\% for small objects (10-20 m). Warning times of the discovered impactors show a similar size dependence: small objects are typically discovered only weeks before impact (median:12.4 days), lower mid-sized within a month (median: 21.5 days), and upper mid-sized objects on timescales of a few months (median: 106.2 days). 39.0\% of large impactors are discovered more than a year before impact, lacking long-lead warning despite their brightness. A loss-mode analysis reveals the underlying cause that small impactors are limited mainly by photometric sensitivity, whereas mid-sized and large objects are missed primarily due to cadence and linking constraints from LSST and its Solar System Processing (SSP) Pipelines. These results show that LSST excels at discovering faint, small impactors, but cannot by itself guarantee long-lead warning across the hazardous size spectrum. Coordinated multi-survey strategies will therefore be essential in the LSST era to achieve robust planetary-defense capability, and we study a complementary high-cadence, shallow-depth example with the Argus Array.

\end{abstract}


\section{Introduction}

Near-Earth Objects (NEOs) evolving onto Earth-crossing trajectories can produce hazardous impacts. Even modest-sized impactors can cause substantial damage, as demonstrated by the 2013 Chelyabinsk airburst ($\sim19-20~\mathrm{m}$) \citep{borovivcka2013trajectory}, which released approximately 500~kt \footnote{$1\,\mathrm{kt\ TNT} = 4.18 \times 10^{12}\,\mathrm{J}$} of energy \citep{brown2013500}. Larger impactors (diameters $\gtrsim 50-100~\mathrm{m}$) pose the potential for regional to global disruption, with objects larger than $\sim140~\mathrm{m}$ representing continental-scale hazards \citep{collins2005earth, mathias2017probabilistic}. These risks place asteroid impacts at the forefront of planetary-defense efforts, and early discovery is essential, as meaningful mitigation and civil-response actions require warning times of months to years \citep{national2010defending}.

The Legacy Survey of Space and Time (LSST), conducted by the Vera C.\ Rubin Observatory (Rubin), is expected to play a major role in future NEO discovery. Recent simulations predict that LSST will discover roughly 53\% of NEOs larger than 140 m over its ten-year baseline \citep{kurlander2025predictions}. However, the same work shows that many NEOs are discovered gradually throughout the survey, rather than concentrated in the early years as for other asteroid populations, indicating that opportunities for long warning times may be missed. Compounding this issue, the rarity of Earth impacts implies that it is challenging with current simulations to include enough impactors to robustly measure LSST's performance for low-probability collision events. Specifically, of the $4.3\times10^{8}$ objects simulated in \citet{kurlander2025predictions}, only $\sim4\times10^3$ (1 out of $10^5$) of these are classified as Potential Hazardous Asteroids (PHAs; minimum orbital intersection distance under 0.05 AU and diameter larger than 140 m), showing the heavy computational cost for analyses of these large simulations. 

Developing a large-scale realistic population of bound Earth-impacting objects suitable for evaluating survey performance remains challenging. Impacting trajectories are intrinsically rare, which has led most prior work to focus on detailed reconstruction of individual impact events \citep[e.g.][]{jenniskens2009impact, brown2013500, farnocchia2016trajectory}. While these case studies provide valuable physical and observational insight for actual impactors, they do not sample the diversity of impact geometries or cadence-driven challenges faced by real survey operations. Survey-wide assessments of NEO detectability have been conducted for systems such as the Asteroid Terrestrial-impact Last Alert System (ATLAS; \citealt{Heinze_2021}) and Panoramic Survey Telescope and Rapid Response System (Pan-STARRS; \citealt{VERES2009472}). However, like the recent LSST simulations, these analyses do not incorporate a dedicated impactor population and therefore are limited in how they probe survey performance on the objects of greatest planetary-defense relevance.

A notable exception is the synthetic Earth-impactor population developed by \citet{Chesley2019}, who generate $\sim3000$ physically realistic, probability-weighted set of bound impactors derived from the de-biased NEO population model of \citet{GRANVIK2018181}. However, this population is restricted to a narrow absolute-magnitude range of $24.75 < H < 25$ (corresponding to $\sim30$--$75$~m for commonly assumed NEO geometric albedos of $p_V \sim 0.04$--0.25) and is computationally expensive to extend: approximately $1.8\times10^{8}$ synthetic NEOs were sampled to produce only 3000 impactors, or roughly one impactor per $\sim6\times10^{4}$ NEO draws. These limitations make this approach computationally expensive to scale across sizes. Recent work presented in IAA Planetary Defense Conference \footnote{See Kiker et al. (2025), IAA Planetary Defense Conference extended abstract: \url{https://iaa.4hdt.ro/event/1/contributions/169/attachments/136/311/IAA-PDC-25-04-42-extabs.pdf}.} have extended this method to larger population, enabling studies of how impact probability and warning time evolve under realistic cadences, with further details to be presented in forthcoming work. 

Population-level studies have also been conducted for interstellar impactors with unbound trajectories \citep{seligman2025distribution}. While scientifically important, these objects represent a distinct and much rarer class of impact events and do not inform the more frequent population of bound near-Earth asteroids that repeatedly traverse the inner Solar System. Overall, there is currently limited synthetic impactor data set within Rubin simulations and across the community to support a robust assessment of LSST’s impactor discovery capability.

To fill the gap with efficient computation, we generate such a population by drawing from the existing synthetic NEO population \citep{kurlander2025predictions} produced by \texttt{NEOMOD3} \citep{NESVORNY2024116110}, which models the intrinsic NEO orbital and size–albedo distributions while accounting for survey selection effects, and then introducing minimal perturbations to produce bound, dynamically consistent Earth-impacting trajectories with high sampling efficiency. This method does not compute or apply collision probability and treats all generated impactors as equally threatening. We then model LSST’s response using the \texttt{Sorcha} survey simulator \citep{merritt2025sorcha}, enabling a time-resolved evaluation of detectability, linking, loss mechanisms, and achievable warning times under the baseline cadence. This framework allows us to determine not only what fraction of impactors LSST discovers and how early, but also why some impactors evade discovery—whether due to geometric inaccessibility, cadence gaps, magnitude limits, or linking failures.  
This analysis is complementary to a corresponding paper by Chow et al. (submitted) which uses a set of actual meter-size and larger impactors, observed as bright meteors in Earth's atmosphere, to estimate LSST's absolute discovery yield and completeness for Earth impactors in the $\sim1-10$ m size range.

Recognizing that the scheduling of LSST is optimized for a broad astrophysical science portfolio rather than a planetary-defense cadence, we explore how complementary facilities such as Argus Array (Argus) \footnote{The Argus Array design and schedule are under active development. The most current specifications and project status are maintained at \url{https://argus.unc.edu/}.} \citep{law2022low}--an all-sky, high-cadence survey with first light expected in 2027-- may recover classes of impactors that LSST systematically misses. Taken together, our LSST impactor simulations and Argus comparison provide the first survey-realistic, population-scale evaluation of LSST’s performance on impactors and offer actionable insight for planetary-defense strategies in the Rubin era.

The paper is organized as follows: in Section~\ref{sec:method}, we describe our method for generating the synthetic impactor population and outline the LSST and \texttt{Sorcha} simulation configuration and analysis setup.   
Section~\ref{sec:results} presents LSST’s impactor discovery performance, quantifies warning times, and analyzes the dominant loss mechanisms.  In Section~\ref{sec:argus}, we describe the configuration and examine the potential complementarity of an Argus-like cadence-rich survey. 
In Section~\ref{sec:discussion}, we evaluate the strengths and limitations of LSST and broader implications for planetary-defense strategies.  
Finally, Section~\ref{sec:conclusions} summarizes the key conclusions.

\section{Methodology}
\label{sec:method}

\subsection{Synthetic Impactor Population}
\begin{figure*}[ht!]
\centering
\includegraphics[width=0.8\textwidth]{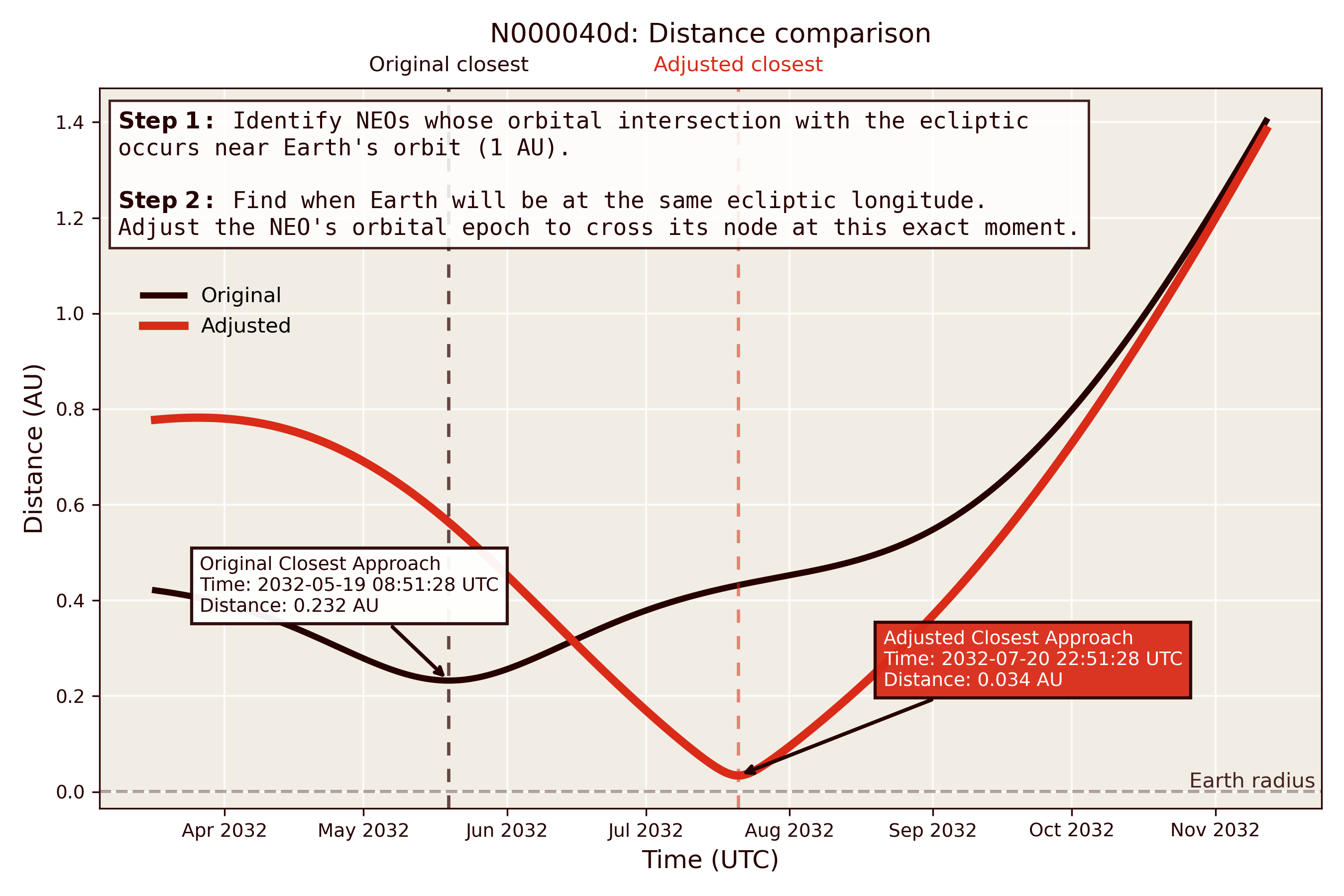}
\caption{Illustration of the synthetic impactor generation model. The plot compares the geocentric distance of a representative NEO before and after epoch adjustment. In Step 1, candidate NEOs are identified with nodal distances at $\sim$1 au. In Step 2, the object’s orbital epoch is shifted so that Earth and the NEO simultaneously arrive at the same ecliptic longitude, forcing an Earth-crossing geometry. The original orbit (black) has a minimum Earth-approach geocentric distance of 0.23 au, while the adjusted orbit (red) produces an impact configuration of closest geocentric approach at 0.03 au. This procedure preserves the object’s orbital elements ($a$, $e$, $i$) while altering the mean anomaly to generate a dynamically consistent synthetic impactor.
\label{fig:impactor_method}}
\end{figure*}

To realistically simulate impactor discoveries, we developed an impact generator that minimally perturbs the orbits of NEOs such that they collide with Earth. This process begins by identifying NEO candidates whose trajectories intersect Earth’s orbit at any time. For each candidate, we record the epoch of this conjunction and compute the ecliptic longitude of the intersection point and Earth’s position at the same epoch. We then calculate the time shift, $t_{\mathrm{shift}}$, needed for Earth to arrive simultaneously at the intersection, defined as
\begin{equation}
    t_{\mathrm{shift}} = t_{\oplus} - t_{\mathrm{ast}}
\end{equation}
where $t_{\oplus}$ denotes the epoch at which Earth reaches the intersection point, and $t_{\mathrm{ast}}$ is the epoch at which the asteroid reaches the same point, i.e., when it crosses their orbital node. 
By adjusting the orbital epoch of the asteroid by $t_{\mathrm{shift}}$, we obtain a synthetic orbit that preserves the dynamical characteristics of the original NEO while ensuring an Earth impact, including the effects of Earth’s gravity. This construction is constrained by requiring the orbit–Earth intersection to occur within 0.04~au of Earth's orbital radius, a criterion chosen to be slightly more restrictive than the standard 0.05~au PHA cutoff while maintaining adequate sample statistics. This approach is illustrated in Figure \ref{fig:impactor_method}.

This approach offers several advantages over synthetic populations constructed from scratch using assumed orbital distributions. It preserves the orbital distributions of inclinations, eccentricities, and semi-major axes seen in the NEO population, ensuring dynamical realism. By conditioning on direct impact, it also provides a sufficiently large and diverse sample of impact geometries at a manageable computational cost, yielding approximately one impactor per eight constructed objects and thereby substantially improving sampling efficiency relative to probability-weighted approaches \citep{Chesley2019}.

We generate 17,424 synthetic impactors from 135,000 parent NEOs to improve statistical robustness while remaining computationally feasible (using a small fraction of the $4.3\times10^8$ total parent NEOs). The impact epochs are drawn uniformly across the LSST survey window, yielding a temporally unbiased set of Earth-impact events for the following detectability assessment. The resulting population spans the characteristic orbital ranges of near-Earth objects: semi-major axes cluster around $1.5$--$1.8$~au, eccentricities peak near $e \sim 0.6$, and inclinations are strongly concentrated below $10^\circ$. The angular elements ($\Omega$, $\omega$, $M$) exhibit nearly uniform coverage, as expected for an unbiased selection from the parent population. The sample is dominated by small objects, with absolute magnitudes $H_r \sim 24$--$27$ (in the $r$ band as provided by the input NEO population) corresponding to diameters of a few tens of meters, and a steep decline toward larger bodies. Together, these properties produce a realistic and dynamically diverse suite of Earth-impacting NEOs appropriate for evaluating survey detectability.

\subsection{Impactor Size Regimes}

Throughout this work, we group impactors into four size regimes: 10–20 m objects producing Chelyabinsk-class airbursts; 20–50 m objects capable of Tunguska-class, city-scale devastation; 50–140 m objects associated with regional-scale impacts; and objects larger than 140 m, corresponding to PHA threshold. For simplicity and clarity in figures and tables, we refer to these regimes as small (10–20 m), lower mid-size (20–50 m), upper mid-size (50–140 m), and large ($>$140 m) impactors.

The boundaries between these regimes are motivated by a combination of impact effects modeling, damage scale and commonly used planetary-defense conventions. Objects in the 10–20 m range correspond to Chelyabinsk-class airbursts, while 20–50 m objects encompass Tunguska-scale events, for which fragmentation models predict qualitatively different breakup behavior and substantially greater air-blast damage when crossing the 20 m threshold \citep{MCMULLAN201919}. Though objects' fragmentation and energy deposition do not exhibit a sharp regime change at larger sizes \citep{WHEELER2017149}, multiple survey-definition efforts and government documents identify $\sim50$ m as a physically and operationally important transition size \citep{NAP26522}. For typical stony objects, this is approximately the lower limit at which atmospheric penetration becomes sufficient to produce significant ground damage from an airburst, and the size above which mitigation and emergency-response considerations become increasingly relevant. Objects larger than 140 m are treated separately, as this threshold corresponds to the planetary-defense community’s definition of PHAs and reflects long-standing policy and hazard-classification conventions. Exact damage outcomes depend on impact angle, velocity, and material properties. Accordingly, these size ranges are intended as approximate, physically motivated categories rather than strict boundaries. 

\subsection{Survey Simulation with Sorcha}
We evaluate the detectability of these synthetic impactors with \texttt{Sorcha}, a survey simulation package developed primarily for LSST Solar System science \citep{merritt2025sorchasolarsurveysimulator, holman2025sorchaoptimizedsolarephemeris}. \texttt{Sorcha} ingests the LSST Operations Simulator (OpSim) outputs, which encode the full survey cadence, field pointings, and limiting magnitudes as a function of time and observing conditions. For each synthetic impactor, Sorcha calculates its apparent magnitudes, sky-plane rates of motion, and observability against LSST’s five-sigma limiting depths and camera footprint of each LSST image. 

All primary analyses, such as discovery statistics and warning-time calculations, use the LSST cadence \texttt{baseline\_v3.4} \footnote{This cadence version is used to maintain consistency with the simulations of \citet{kurlander2025predictions}. A more recent cadence version (\texttt{v5.1}) is publicly available at \url{https://github.com/lsst-sims/sims_featureScheduler_runs5.1}.} with a median single-visit depth of $m_r \simeq 23.95$, and a total exposure time of 30~s per visit, together with the full set of \texttt{Sorcha} internal filtering including trailing-loss corrections, bright-limit filtering, photometric randomization, the fading function and the default linking requirement of at least two detections per night on at least three nights within 15 days \citep{Ivezic2019}. The parameter setup is listed in Table~\ref{tab:lsst_sim_summary}. 

\begin{table*}[t]
\centering
\caption{Summary of LSST simulation inputs used in the Sorcha survey modeling. 
The full configuration shown here is the default setup used for discovery and warning-time analyses.}
\label{tab:lsst_sim_summary}
\begin{tabular}{lcc}
\hline\hline
\textbf{Parameter / Filter} & \textbf{Symbol / Setting} & \textbf{Value / Source} \\
\hline
\multicolumn{3}{c}{\textit{Survey Cadence and Depth (Baseline Pointing File)}} \\
\hline
Cadence baseline                 &                & \texttt{baseline\_v3.4} OpSim \\
Single-visit depth ($r$ band)    & $m_r$          & Median 23.95 mag \\
Exposure time per visit          & $t_{\mathrm{exp}}$ & 30 s \\
\hline
\multicolumn{3}{c}{\textit{Full Sorcha Filtering, Noise, and Linking Models}} \\
\hline
Bright-limit filtering     &  & 16.0\\
Trailing loss correction    &  & ON \\
Vignetting model              &  & ON \\
Photometric randomization &  & ON \\
\hline
\multirow{3}{*}{Fading function}
 & Width & 0.1\\
 & Peak efficiency & 1\\
\hline
\multirow{5}{*}{Linking/tracklet requirement}
 & Detections per tracklet & $\geq$2 per night \\
 & Minimum intra-night separation & 0.5 arcsec \\
 & Maximum intra-night time span & 0.0625 days (90 min) \\
 & Tracklets per discovery & $\geq$3 \\
 & Tracklet time window & 15 days \\
\hline

\hline
\end{tabular}
\end{table*}

An object is identified as ``discovered" only if its detections or observations pass the full filtering and successfully satisfies the linkage criterion. Appendix~\ref{appendix:sorcha_configuration} provides the detailed, stepwise \texttt{Sorcha} configurations used to isolate each survey-performance component.

LSST survey simulations were executed on Duke Computer Cluster (DCC). With the ten-year \texttt{baseline\_v3.4} pointing database, we ran the full synthetic impactor population simulation as a single-node, high-memory job (300–400 GB RAM), with typical wall-clock runtime of 4–5 hours. 

\subsection{Detectability Diagnostics and Classification}
\label{sec:methods-detection-categories}
We compute a set of detectability metrics designed to quantify both the survey's sensitivity and the reasons individual objects are missed. These include (1) the discovery efficiency as a function of impactor diameter; (2) the pre-impact warning time, defined as $t_{\rm impact}-t_{\rm first\,link}$, where $t_{\rm first\,link}$ is the \texttt{date\_linked\_MJD} from Sorcha output, specifying epoch where objects are discovered by forming successful linkage; (3) the apparent magnitude, rate of motion, and solar elongation at discovery; and (4) the orbital and geometric properties of discovered objects. 

\begin{deluxetable*}{lccc}
\tablecaption{Outcome Categories Used to Classify Missed and Discovered Impactors\label{tab:miss_categories}}
\tablehead{
\colhead{Category}&
\colhead{Ever in FOV} &
\colhead{Any Detections} &
\colhead{Linked Detections}
}
\startdata
\textbf{Pointing loss}            & No  & No  & No  \\
\textbf{Photometric loss} & Yes  & No  & No \\
\textbf{Linking loss}   & Yes & Yes  & No \\
\textbf{Linked \& discovered}       & Yes & Yes & Yes \\
\enddata
\tablecomments{Each object falls into exactly one category. Definitions are described in Section~\ref{sec:methods-detection-categories}.}
\end{deluxetable*}


To interpret these diagnostics, each impactor is assigned to one of four mutually exclusive outcome categories (Table~\ref{tab:miss_categories}). These categories disentangle the distinct mechanisms by which LSST fails to discover an impactor: \emph{pointing losses} identify objects never placed within the field of view (FOV); \emph{photometric losses} isolate cases where the object enters the FOV but produces no viable observations; and \emph{linking losses} capture objects that are bright enough and observed at least once but not revisited with sufficient cadence to form a linkage. Objects with successful detections and linking are classified as \emph{linked and discovered}. This framework allows us to separate geometric limitations, photometric constraints, and cadence-driven failures, enabling a physically interpretable accounting of LSST's missed impact population.

\section{LSST Impactor Discovery Results}
\label{sec:results}

\subsection{Discovery Rates}

The overall discovery efficiency of LSST across our synthetic impactor population is 13.7\%, with performance strongly dependent on object size. For objects larger than 140 m, the discovery rate rises to 79.7\%, comparable to NASA’s stated 90\% completness goal for this regime \citep{USCongress2005NEO}. At smaller sizes, however, discovery efficiency decreases sharply: only 10.5\% of 10--20 m objects, and 26.8\% of objects in the 20--50 m Tunguska-scale range are discovered, and fewer than 6\% of Chelyabinsk-scale ($<20$ m) objects are identified prior to impact. This last number is consistent with the $\sim$4.1\% pre-impact discovery fraction reported by Chow et al. (submitted) for a set of 1–10 m actual Earth impactors.

\begin{deluxetable}{lrrr}
\tablecaption{Discovery efficiency by size (LSST)\label{tab:detection_varbins}}
\tablewidth{0pt}
\tablehead{
\colhead{Size Range (m)} & \colhead{All} & \colhead{Discovered} & \colhead{Discovery Rate}
}
\startdata
(10, 20) & 14419 & 1509 & 0.105 \\
(20, 50) & 2747 & 735 & 0.268 \\
(50, 140) & 199 & 100 & 0.503 \\
(140, 2210) & 59 & 47 & 0.797 \\
\hline
Overall & 17424 & 2391 & 0.137 \\
\enddata
\tablecomments{Discovery rate = Discovered / All per bin. Size bins are chosen to reflect impact-hazard regimes: 10--20 m (Chelyabinsk-scale airbursts), 20--50 m
(Tunguska-scale city-level impacts), 50--140 m (regional-scale impacts), and
$>140$ m (the PHA threshold). The final row reports the overall discovery rate
across the full synthetic population.}
\end{deluxetable}

\begin{figure*}[ht!]
\centering
\includegraphics[width=0.8\textwidth]{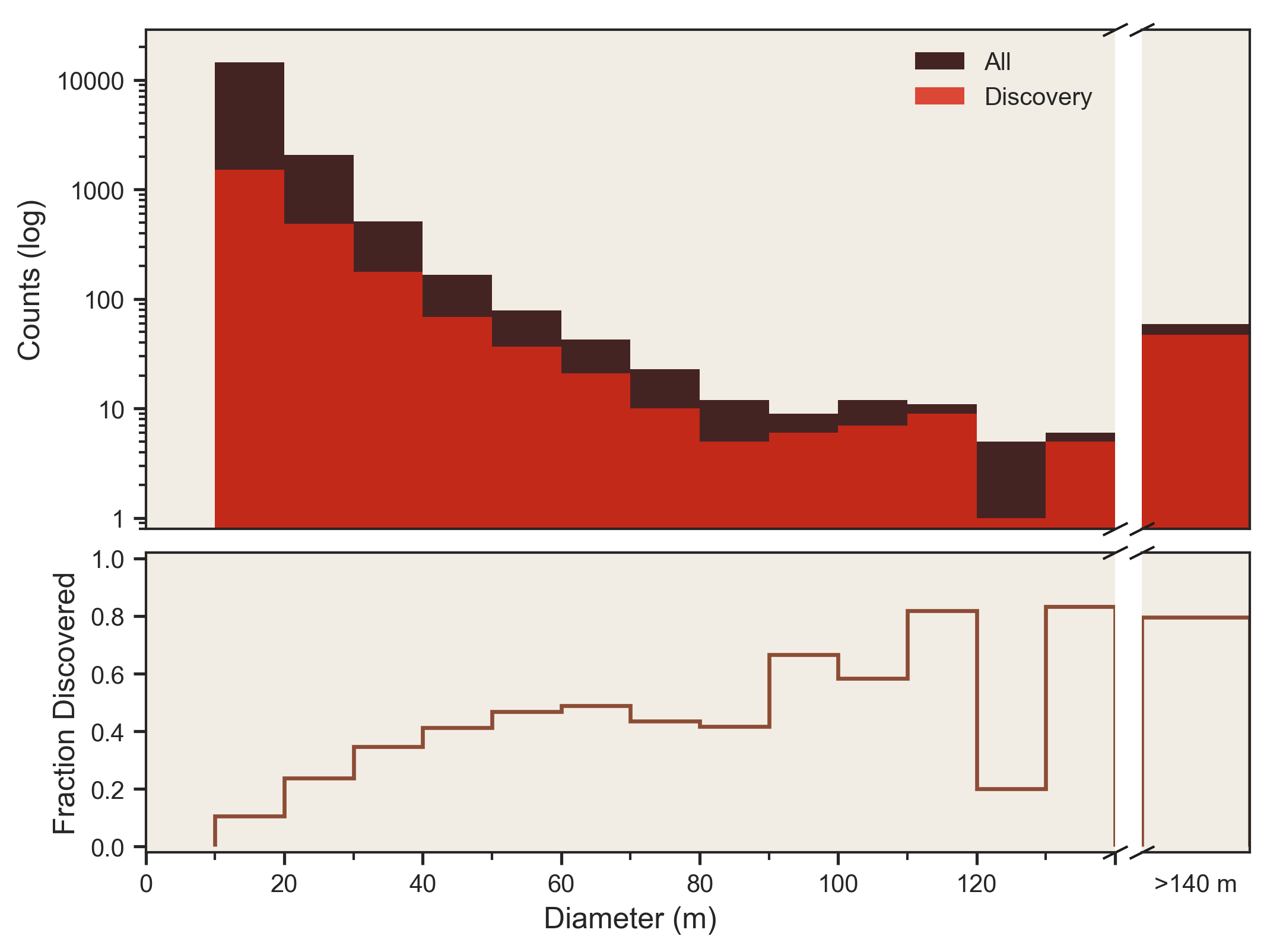}
\caption{
Size distribution and discovery efficiency of the synthetic impactor population, computed using uniform 10 m diameter bins up to 140 m and a collapsed terminal bin for all objects larger than 140 m. 
\emph{Top:} Log-scaled histograms showing the full population (dark brown) and the subset discovered by LSST (red). A break in the x-axis isolates the $>140$ m population corresponding to the PHA hazard regime.
\emph{Bottom:} Discovery fraction (Discovered/All) for the same size-bin setup. 
\label{fig:det_rate_by_size}}

\end{figure*}

Table~\ref{tab:detection_varbins} summarizes discovery efficiency across the major hazard-relevant size regimes, while Figure~\ref{fig:det_rate_by_size} shows the detailed per-bin discovery fraction using uniform 10 m diameter bins (with a collapsed terminal bin for $>140$ m). The histogram trend illustrates a systematic rise in discovery probability with size, reflecting the combined effects of increased intrinsic brightness and a longer time window during which these objects remain above LSST’s limiting magnitude. The efficiency begins to climb steadily above $\sim$20 m, shows modest fluctuations between 50--140 m where population statistics are sparse, and reaches $\sim80\%$ for objects larger than 140 m. This is broadly consistent with the results of Kiker et al., who report a discovery fraction of 70.4\% for $>140$ m impactors, with the difference reflecting their inclusion of impactors with impact dates extending beyond the LSST survey window. Overall, the results highlight the strong size dependence of pre-impact detectability and the challenges posed by small, fast-approaching impactors.

\subsection{Warning Times}

Early discovery is a central requirement for an effective planetary-defense system, making the pre-impact warning time a critical performance metric. To facilitate interpretation, we present warning-time results both for the full synthetic impactor population and grouped by object size. These warning times are conditioned on impacts occurring during the LSST survey window. 

\begin{deluxetable*}{lrrccccccc}
\tablecaption{Warning-time statistics by impactor size (LSST). \label{tab:size_warning_lsst}}
\tablehead{
\colhead{Size bin (m)} &\colhead{Median$^\dagger$} &\colhead{P10$^\dagger$} &\colhead{P90$^\dagger$} &\colhead{$<1$ day} &\colhead{$1$--$7$ days} &\colhead{$1$ week--$1$ month} &\colhead{$1$--$6$ months} &\colhead{$6$ months--$1$ year} &\colhead{$>1$ year}
}
\startdata
10–20 m & 12.4 & 3.3 & 45.9 &0.1\% & 1.4\% & 3.1\% & 0.5\% & 0.0\% & 0.4\% \\ 
20–50 m & 21.5 & 5.0 & 1343.2 &0.1\% & 1.9\% & 6.6\% & 3.1\% & 0.2\% & 2.1\% \\ 
50–140 m & 106.2 & 18.5 & 1884.1 &0.0\% & 2.0\% & 5.0\% & 14.1\% & 1.5\% & 10.1\% \\ 
$>$140 m & 1218.4 & 56.9 & 2952.3 &0.0\% & 1.7\% & 3.4\% & 18.6\% & 1.7\% & 39.0\% \\ 
\enddata
\tablecomments{Columns labeled by thresholds report the fraction of the \emph{full} impactor population in each size bin discovered at least that long before impact (i.e., $P(W\ge X)$). Warning-time quantiles ($^\dagger$) are computed only for discovered objects with valid warning times.}
\end{deluxetable*}

\begin{figure*}[ht!]
\centering
\includegraphics[width=0.7\textwidth]{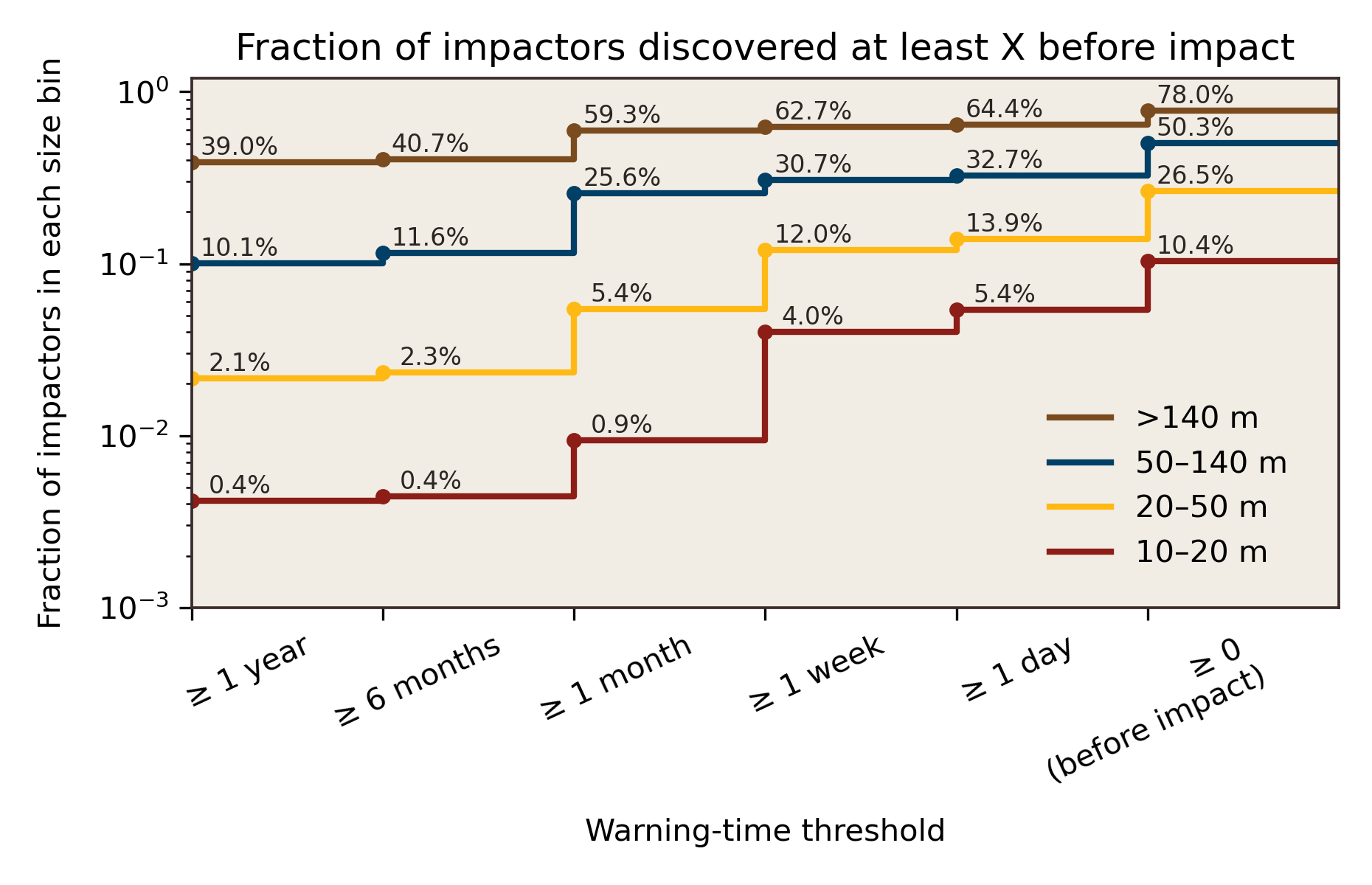}
\caption{
Fraction of impactors discovered at least $X$ (time) before impact, shown for four size bins (10--20~m, 20--50~m, 50--140~m, and $>140$~m). Values represent cumulative fractions of the full impactor population in each size bin discovered with warning times $\ge X$; the $\ge 0$ bin corresponds to full discovery efficiency. The y axis is logarithmic. Discovery probability and warning time increase strongly with impactor size, with small objects rarely discovered far in advance.
Most discovered impactors receive warning times of only weeks to months, while year-scale warning is rare.
\label{fig:warning_lsst}}
\end{figure*}

Across all sizes, LSST discoveries are strongly concentrated at short notice. 5.76\% of the full population are discovered at least one week before impact, dropping to 0.91\% receiving more than one year of warning. This indicates that, although LSST’s depth enables long-lead observations for a minority of favorable geometries, most impactors are discovered too late for mitigation options beyond civil defense. 

A strong size dependence emerges when warning times are binned by diameter. Figure~\ref{fig:warning_lsst} shows the cumulative fraction of impactors discovered as a function of time before impact for the four size bins, with corresponding values in Tables~\ref{tab:size_warning_lsst}. Large ($>$140 m) impactors are typically discovered months to years before impact, with a median warning time of 1218.4 days and a broad 10th–90th percentile span of a few months to a few years (56.9 - 2953.2 days). More than 40\% of these impactors receive $>$6 months of warning. For upper mid-size objects (50--140 m), the median warning time drops sharply to 106.2 days, and only 11.6\% are discovered $\le$6 months before impact. Nearly 21.1\% receive less than six months of warning, including 7.0\% discovered within one month. The small and lower mid-sized objects (10--50 m) experience the most severe warning-time limitations. Their median warning time are both less than a month, with a wide tail but a large fraction discovered late. Less than 3\% of these impactors receive more than six months of warning, despite occasional long-tail discoveries in favorable geometries.

Overall, LSST’s warning-time distribution reveals a steep transition between large impactors discovered with ample warning, and the 10--50 m class, for which days to weeks of lead time is typical. At even smaller sizes (1–10 m), Chow et al. (submitted) find a median time of discovery of only $1.70$ days before impact for impactors discovered with LSST. 

\subsection{Early-survey impact-time bias.}
Impactors occurring during the earliest phase of the survey experience an intrinsically truncated discovery window, leading to shorter achievable warning times independent of survey performance. In our synthetic population, impact epochs are uniformly distributed over the survey duration, implying that only a small fraction of objects impact shortly after survey start and limiting the potential contribution of this effect to $\lesssim 10\%$ of the total sample. 

To explicitly quantify this bias, we repeat the discovery-rate and warning-time analysis after excluding all impactors with impact dates prior to 2027 May 1, corresponding to approximately the first two years of survey operations. 

Removing these early impacts produces only modest changes in discovery efficiency across all size bins (at the level of a few percent) and does not alter the size-dependent discovery trends reported in the main analysis. Median warning times increase slightly for larger objects, reflecting the removal of intrinsically time-limited early impacts rather than any change in survey sensitivity. Full quantitative results for this test are presented in Appendix~\ref{appendix:early_bias}.

\subsection{Discovery Circumstances}
\label{subsec:discovery_circumstances}
The circumstances of the first detection determine how much time and orbital information are available for subsequent impact assessment. For each simulated impactor that is discovered prior to impact, we identify the earliest LSST observation and characterize its geometric, kinematic, and photometric properties.  

Geometrically, LSST tends to first detect impactors when they are already near Earth, mostoccurring at small geocentric distances ($\Delta \lesssim 0.1$~AU) and heliocentric distances near Earth’s orbit. Smaller impactors are generally discovered only at very close approach, whereas larger objects are first detected farther out, reflecting a strong size dependence. Discoveries overwhelmingly occur at large solar elongations near opposition, but at moderate phase angles, so objects are only partially illuminated. At these epochs, impactors usually exhibit modest sky-plane rates and low transverse velocities relative to Earth, indicating motion largely along the line of sight rather than rapid angular sweeping. Photometrically, first detections cluster within $\sim$1–2 magnitudes of the single-visit depth, with signal-to-noise ratios above threshold and trail lengths smaller than the seeing disk. The detailed analysis on first detection geometry, kinematics, and photometry are shown in Appendix~\ref{fig:appendix_discovery_diagnostics}. Together, these trends suggest that the opportunity for pre-impact discovery is shaped by a combination of viewing geometry and survey depth.

\subsection{Classification of Missed Objects and LSST Survey Constraints}
\label{sec:missed_classification}

To identify the dominant survey limitations affecting impactor discovery, we trace every synthetic impactor through its observational history summarized in Table~\ref{tab:impactor_loss_summary_lsst}. 
\begin{table*}
\centering
\caption{Loss mode classification (LSST)}
\label{tab:impactor_loss_summary_lsst}
\begin{tabular}{llrrrr}
\hline
Category & Subcategory & $N_{\mathrm{obj}}$ & Fraction of all & $N_{\mathrm{missed}}$ & Fraction of missed \\
\hline
\multirow{2}{*}{Cadence loss}
 & Pointing loss & 5    & 0.000 & 5    & 0.000 \\
 & Linking loss  & 5133 & 0.295 & 5133 & 0.341 \\
\hline
Photometric loss & & 9895 & 0.568 & 9895 & 0.658 \\
\hline
Linked and discovered & & 2391 & 0.137 & -- & -- \\
\hline
\end{tabular}
\end{table*}

Pointing losses are negligible ($<0.001$) and occur primarily for impactors whose trajectories lie in the northern hemisphere, indicating that Earth-impacting trajectories are almost always caught by the Rubin Observatory field of view at least once under the adopted survey cadence. In contrast, magnitude or trailing loss accounts for 65.8\% of all missed objects and represent the dominant sources of LSST incompleteness. These losses occur primarily for the smallest impactors, whose apparent brightness remains fainter than $r\!\sim\!24.5$ even during close approaches, or whose trails reduce discovery efficiency. The other 34.1\% cases are lost at linking stage. These failures arise from seasonal gaps, weather outages, sparse revisit cadence, and field coverage patterns that prevent the formation of linkable tracklets.

For 10--20 m objects, photometric losses dominate ($\approx 69\%$), while linking loss accounts for $\approx 31\%$ of missed impactors in this bin. The two mechanisms are roughly equal for 20--50~m impactors ($\approx 50\%$ each). At larger sizes, linking losses increasingly dominate, comprising $\approx 74\%$ of missed objects at 50--140~m and $\approx 83\%$ above 140 m, indicating a shift from sensitivity-limited losses at small sizes to linking-limited losses for larger impactors. 

Overall, LSST’s incompleteness is not driven by geometric reach of camera footprint given that nearly all impactors enter the survey footprint at some point. Instead, the dominant limitations arise from (i) photometric losses for small objects and (ii) cadence-driven linking gaps for moderate- and large-sized impactors.  

This population-level diagnosis motivates the detailed analysis in the following subsection, where we examine through both population filtering and individual case studies how cadence and linking constraints shape LSST’s impactor discovery performance.

\subsection{Cadence and Linking Effects}
\label{sec:cadence_linking}

The discovery rate and loss-mode analyses in the previous sections established that cadence-driven linking failures are the dominant source of incompleteness for impactors that LSST does observe at least once. Here we examine the underlying mechanisms that produce these losses and demonstrate how visibility, detectability and temporal sampling combine to determine whether an object is ultimately discovered. 

At population level, Appendix~\ref{fig:panel_lsst_stages} shows how each Sorcha stage removes different subsets of the synthetic population. The first stages -- magnitude limits, randomization, and the fading function -- primarily suppress the smallest objects, consistent with their expected photometric limitations. By contrast, the steepest drop in completeness for moderate-sized ($\sim$50--300 m) impactors occurs at the linking stage. 

\begin{figure*}[htp!]
\centering

\gridline{
  \fig{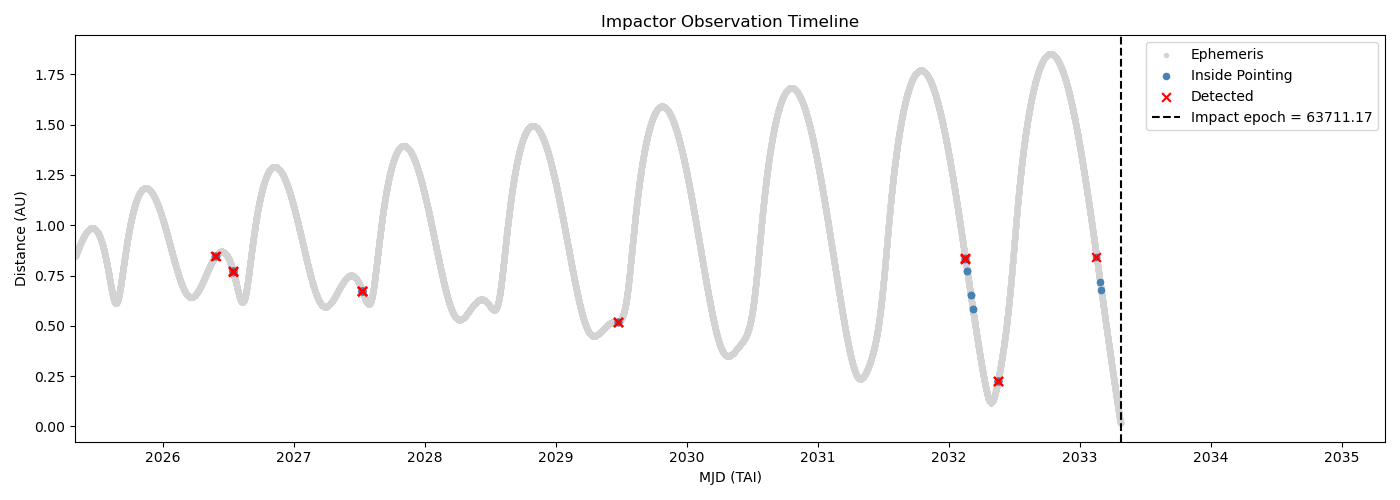}{0.95\textwidth}{
  \begin{minipage}{\linewidth}
  \raggedright
  (a) \textbf{Unlinked impactor with partial visibility.}
  The gray curve shows the geocentric distance as a function of time. 
  Grey points trace the full ephemeris; blue points mark epochs when the object falls within LSST’s pointing footprint. The absence of blue points along large portions of the orbit reflects pure \emph{pointing loss}, where the survey never visits the object’s location. Where blue points do appear, many lack over-plotted red crosses, indicating that the object is inside the footprint but remains below the single-visit magnitude limit or suffers trailing losses—i.e.\ \emph{magnitude loss}. Even the epochs with successful synthetic detections (red crosses) occur too sparsely in time to form intra-night or inter-night pairs, preventing any tracklet formation. The result is \emph{linking failure} despite clear visibility windows.
  \end{minipage}
}
}
\vspace{-0.6em}
\gridline{
  \fig{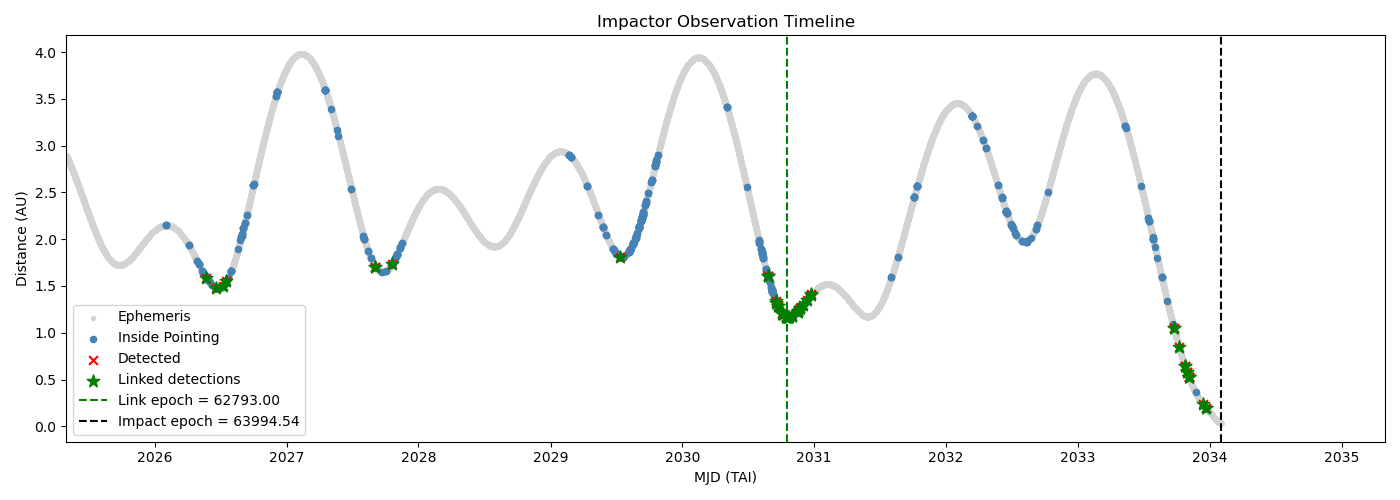}{0.95\textwidth}{
  \begin{minipage}{\linewidth}
  \raggedright
  (b) \textbf{Successful linked detections.}
  A contrasting case where the object repeatedly enters LSST’s pointing footprint (blue points) at times when it is also bright enough to be detected (red crosses). These detections occur with the cadence required to form valid intra-night tracklets and inter-night linkages. Green star indicate the subset of detections incorporated into the final linked solution, and the dashed vertical line marks the epoch of first linkage. This case illustrates that successful discovery requires the coincidence of three conditions: pointing coverage, sufficient brightness, and adequate temporal sampling for linking.
  \end{minipage}
  }
}

\caption{
Two representative synthetic-impact cases illustrating how visibility does—or does not—translate into discovery.  
Panel (a) demonstrates all three loss mechanisms: segments with no blue points represent pointing loss; blue points without red crosses indicate magnitude or trailing loss; and isolated red-cross detections that never form tracklets represent linking loss.  
Panel (b) shows a case where all conditions align, enabling full linkage and discovery. Together these two examples encapsulate the essential survey failure modes that govern LSST's impactor completeness.}
\label{fig:visibility_vs_discovery}
\end{figure*}

At individual level, Figure~\ref{fig:visibility_vs_discovery} illustrates how these loss mechanisms manifest, exemplified by two contrasting cases. These examples reinforce the population-level result that LSST's discovery success hinges not only on visibility or brightness but on whether observations arrive with the temporal structure required for tracklet formation.

\subsection{LSST Strengths and Limitations}

LSST’s impactor performance is governed by three requirements: the object must enter the pointing footprint, be brighter than the detection threshold, and receive temporally clustered visits that allow three tracklets formation. LSST excels when these conditions hold over long timescales—its depth and sky coverage yield early detections of large, faint impactors that few surveys can access.

For small or rapidly approaching objects, however, incompleteness is driven not by sensitivity but by cadence and linking strategies. Many impactors are either geometrically inaccessible or observed only in isolated single visits that cannot be linked. Warning-time statistics reflect this: large ($>140$ m) objects are found months to years in advance, while smaller impactors receive sharply reduced lead times due to their limited visibility during each close approach. LSST often sees these objects, but not with sufficient number of observations in a linkable temporal pattern.

\section{Characterizing the Argus Survey Performances}
\label{sec:argus}

The cadence driven limitation from LSST motivates our examination of whether high-cadence systems—such as Argus-like all-sky arrays—can serve as an complementary resource. Designed for continuous all-sky monitoring, Argus is expected to cover $\sim$21,000~deg$^{2}$ \citep{law2022low} per night with minute-scale cadence in dark and gray time (and sub-minute cadence in bright time), providing near-continuous temporal sampling for fast-moving and short-warning-time impactors.
A system with such high cadence may recover impactors that LSST detects but cannot link, allowing it to complement LSST in discovering impactors. However, Argus Array's comparatively shallow depth \citep[$m_g \sim 19.6$]{law2022low} compared to LSST may limit its performance, motivating a quantitative assessment.

\subsection{Argus simulation setup and conservative performance assumptions}

Given that the moving object pipeline design for Argus has not been finalized, we conducted limited tests by pairing Argus-like survey characteristics, such as single-exposure depth, sky coverage and continuous cadence \citep{law2022low, corbett2022skyterabitsecondarchitecture}, with LSST's Moving Object Processing System (MOPS) implemented in \texttt{Sorcha}. The resulting simulated survey serves as a testbed for isolating the influence of uninterrupted temporal coverage relative to LSST’s sparse revisit pattern with realistic setting, rather than as a predictive forecast of final Argus performance. 

The assumed pointing strategy follows the local zenith of a fixed Earth location throughout each night, with solar altitude $<0^\circ$, consistent with the current Argus operation setting. This produces a natural full night sky sweep as the Earth rotates during dark time. Each field is observed with 30~s exposures, yielding a sub-minute revisit cadence across the visible hemisphere.

For photometric performance, we adopt a conservative single-visit limiting magnitude for the Argus system, assuming an effective $r$-band depth of $m_r \sim 19.6$ as a simplified proxy for the $g$-band performance ($m_g \sim 19.6$; see \cite{law2022low}), and we do not enable frame co-addition, multi-site image combination, or depth-enhancing post-processing. In practice, the Argus Array is expected to co-add both spatially and temporally to achieve deeper effective limits. 

To enable one-to-one comparison with LSST, the Argus-like survey is processed through the same four-stage \texttt{Sorcha} filtering sequence including vignetting and magnitude-limit cuts, per-visit photometric randomization, fading-function efficiency roll-off near the limiting magnitude, and the LSST-style tracklet-linking requirement. We also apply the same analysis to evaluate discovery rate, warning time, loss mode, linking behavior. Discovery for Argus refers to objects that are successfully linked by the LSST MOPS. 

Due to the large cadence file, the Argus simulations were executed on the DCC using a 23-element job array, with the synthetic impactor population divided into batches of 768 objects per job, internally processed using parallel workers to keep per-job memory usage below $\sim$700~GB. Each job used up to 32 CPU cores and typically required $\sim$5 hours of wall-clock time. The outputs were merged to form the final Argus survey simulation results.

\subsection{Argus discovery efficiency}

The discovery efficiency of Argus rises with impactor size, reflecting the shallow single-visit depth adopted in our simulation. As summarized in Table~\ref{tab:detection_varbins_argus}, completeness is very low for objects smaller than $\sim 50$ m, reaching only $0.2$--$0.9$\% across the 10--50 m range. Only in the largest bin ($>140$ m), where objects remain bright for longer fractions of their approach and exhibit slower photometric evolution, does Argus achieve moderate completeness ($\sim 61$\%). Detailed discovery efficiencies binned in 10 m intervals analogous to the LSST analysis are shown in Appendix~\ref{appendix: stage_filtering}, indicating that Argus is intrinsically limited in recovering small, low brightness objects.

\begin{deluxetable}{lrrr}
\tablecaption{Discovery efficiency by size (Argus) \label{tab:detection_varbins_argus}}
\tablewidth{0pt}
\tablehead{
\colhead{Size Range (m)} & \colhead{All} & \colhead{Discovered} & \colhead{Discovery Rate}
}
\startdata
(10, 20) & 14419 & 28 & 0.002 \\
(20, 50) & 2747 & 25 & 0.009 \\
(50, 140) & 199 & 20 & 0.101 \\
(140, 2210) & 59 & 36 & 0.610 \\
\hline
Overall & 17424 & 109 & 0.006 \\
\enddata
\tablecomments{Discovery rate = Discovered/All per bin.}
\end{deluxetable}

\subsection{Warning-time distributions: LSST versus Argus}

\begin{figure*}[ht!]
\centering
\includegraphics[width=0.85\textwidth]{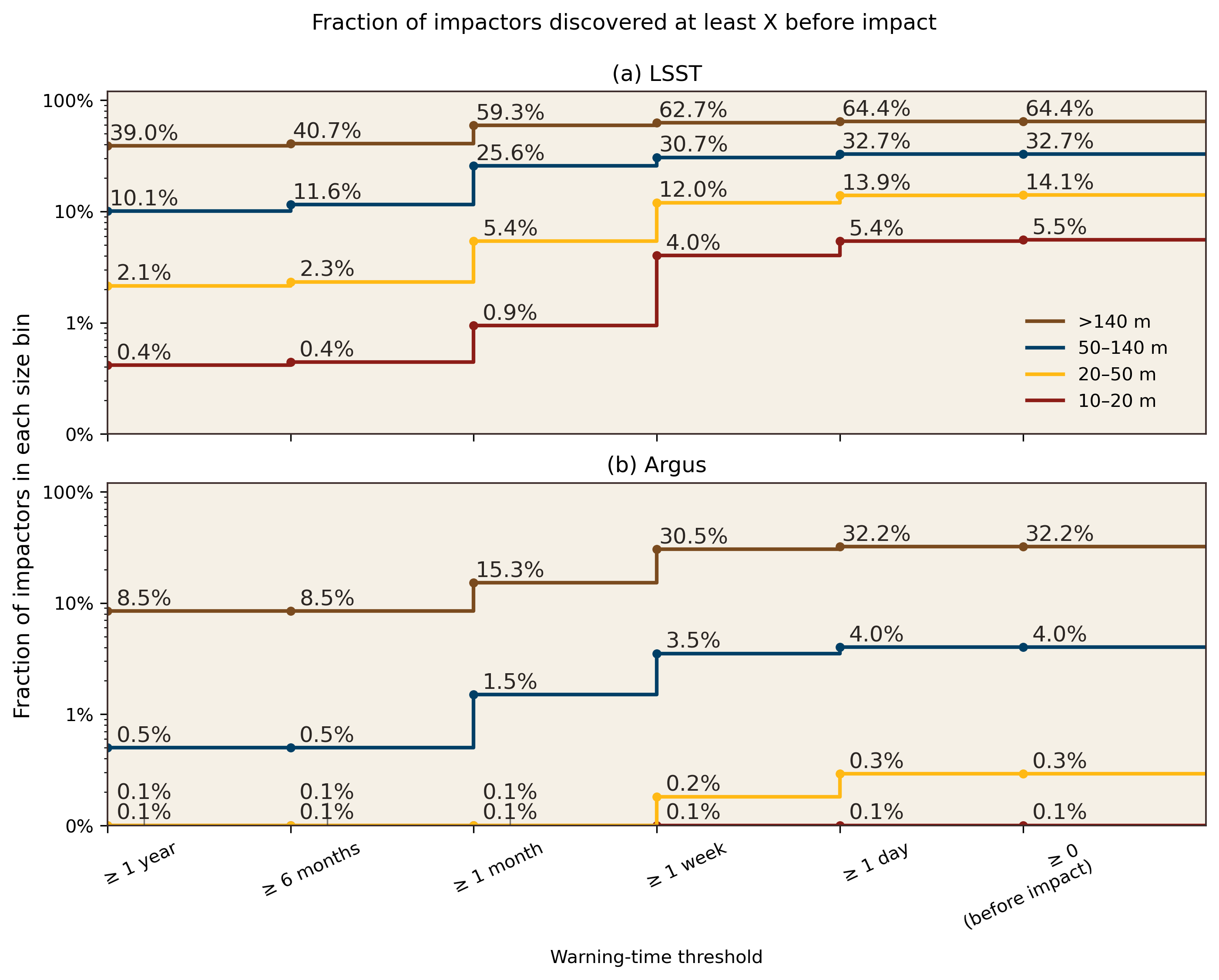}
\caption{
Fraction of impactors discovered at least X before impact for four diameter bins (10–20 m, 20–50 m, 50–140 m, and $>$140 m), shown for (a) LSST and (b) an Argus-like survey. The y-axis lower limit is clipped at \(10^{-3}\) (0.1\%) to avoid unnecessary blank space. For the Argus case, several size--time combinations fall at or below this level, indicating discovery fractions consistent with close to zero at the resolution of this plot.
}.
\label{fig:violin_size_warning_argus}
\end{figure*}

The majority of Argus discoveries occur within a few weeks of impact across all sizes, with nearly half taking place in the final week. For small and lower mid-sized impactors ($10 - 50$ m), almost all are discovered within a week or even on the same day of impact, and nearly none are identified on multi-month timescales. Upper mid-sized objects (50--140 m) exhibit somewhat longer visibility intervals, but still yield predominantly short-warning discoveries, with only a small fraction (0.5\%) exceeding six months of lead time. Even the $>140$\,m population, which LSST typically discovers months or years in advance, shows a substantially shortened warning-time distribution under Argus-like depth assumptions. A comparison of warning time distribution of discovered objects between LSST and Argus are shown in Figure~\ref{fig:violin_size_warning_argus}.

This systematic shift toward short-warning discoveries arises mainly from the survey’s limited photometric reach, which is further discussed in section~\ref{sec:LSST_and_argus_discussion}

\subsection{Argus Loss Mode Analysis}

\begin{table*}
\centering
\caption{Loss mode classification (Argus)}
\label{tab:argus_loss_summary}
\begin{tabular}{lrrrr}
\hline
Category & $N$ & Fraction of all & $N_{\rm missed}$ & Fraction of missed \\
\hline
Photometric loss & 17312 & 0.994 & 17312 & 0.998 \\
Pointing loss & 3 & 0.000 & 3 & 0.000 \\
Linking loss & 35 & 0.002 & 35 & 0.002 \\
Linked discovered & 74 & 0.004 & - & - \\

\hline
\end{tabular}
\end{table*}
An analogous loss-mode analysis for the Argus is summarized in Table~\ref{tab:argus_loss_summary}. Despite continuous monitoring, a very small number (3 out of 17,424) of impactors are still lost due to pointing constraints. Examination of these cases shows that they correspond to objects spending the majority of their time in the Southern sky, with a brief visibility in the northern sky occurring during local daytime at the Argus Array site, similar to the pointing-loss mechanism seen for LSST. The dominant source of incompleteness arises from magnitude losses, which account for nearly all missed impactors. The ratio of linking losses to successfully linked discoveries (0.47:1) indicates that linking plays a much smaller role in Argus incompleteness than in the LSST results (2.14:1). Stage-by-stage filtering (Appendix~\ref{fig:argus_stages}) confirms that Argus losses are almost entirely photometric, with linking failures playing a negligible role once objects exceed the detection threshold. However, the linking requirement introduces a noticeable decline in the discovery rate around 120 - 130~m, similar to that seen in LSST result.

\section{Discussion}
\label{sec:discussion}

\subsection{Complementary Role Between LSST and Argus}
\label{sec:LSST_and_argus_discussion}
The comparison results of LSST and Argus show that the two systems succeed and fail for fundamentally different reasons. LSST's depth enables the early discovery of faint, distance impactors, but its sparse revisit spacing leaves many of these objects unlinked, particularly for impactor sizes most relevant to regional-scale hazards (30--100 m). In contrast, Argus is shallow, eliminating most impactors in the photometric stages, yet any object that does brighten above its limit is observed in dense, continuous sequences. 

\begin{figure*}[ht!]
\centering
\includegraphics[width=1\textwidth]{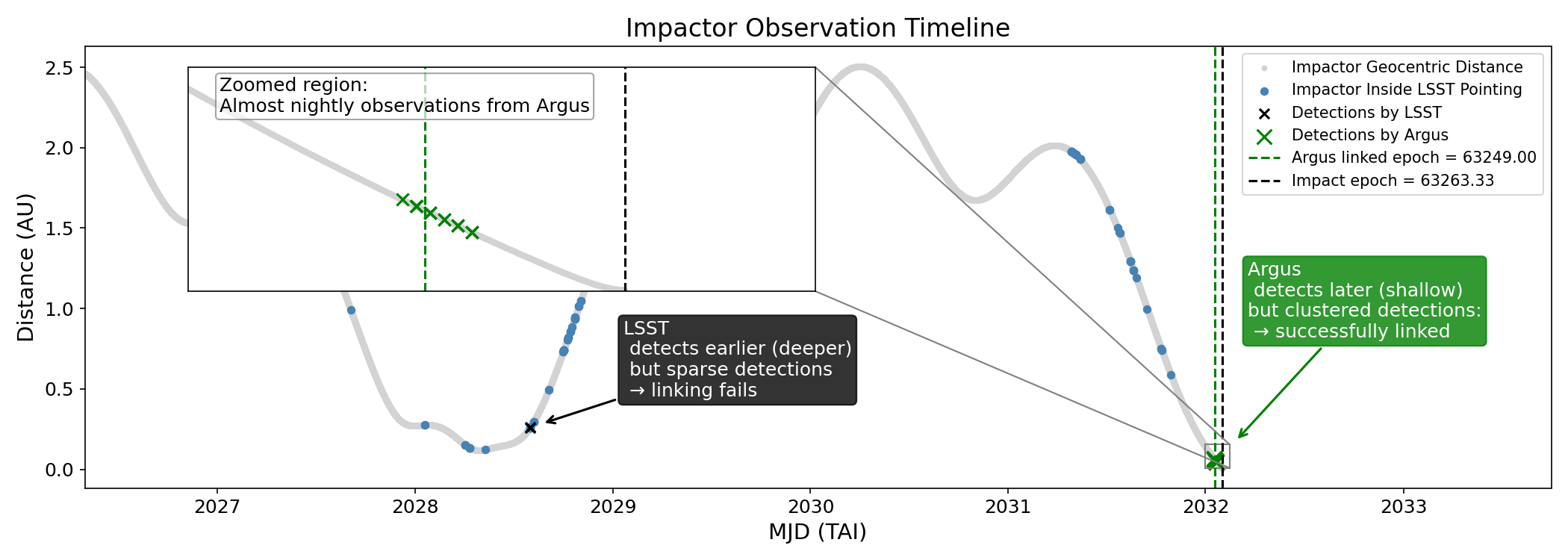}
\caption{Example impactor observation timeline illustrating a case where LSST detects the object but fails to achieve linkage, while Argus successfully links it. Grey curves show the object’s geocentric distance, with blue points marking times when it falls within LSST’s footprint. LSST detections (black crosses) are too sparsely spaced to satisfy linking requirements, whereas Argus detections (green crosses) form dense nightly clusters near close approach, enabling robust linkage, as highlighted in the inset. Vertical dashed lines show the Argus link epoch (green) and the impact epoch (black).
\label{fig:argus_individual}}
\end{figure*}

For detectable objects, Argus produces orders of magnitude more observations per apparition than LSST, yielding immediate linkage even at close approach. Figure~\ref{fig:argus_individual} illustrates this complementarity: LSST detects the object earlier but too sparsely to link, whereas Argus detects it later but with cadence sufficient for instantaneous linkage. Argus therefore recovers impactors that LSST misses due to temporal under-sampling, particularly fast-moving or rapidly brightening objects with short-lived apparitions. Under our assumptions, Argus improves completeness mainly in the bright, late-warning regime, and its reach could expand with modest temporal stacking. 

More broadly, these results demonstrate the value of pairing LSST with an even higher-cadence survey. A cadence-rich system such as Argus mitigates LSST’s temporal losses while LSST retains its unique capability for early, faint detections. Together, they can potentially form a more robust impactor discovery network than either system can achieve alone.

\subsection{Limitations and Future Work}

While our simulations provide a comprehensive assessment of LSST's impactor detectability and highlight the complementary strengths of an Argus-like survey, several methodological limitations exist that motivate future work.

\begin{itemize}
    \item \textit{Synthetic impactor generation:}
    Our synthetic population enforces close Earth encounters within a geocentric distance of 0.04~au but does not require an exact Earth impact. Although this ensures a dynamically representative set of imminent hazard cases, it omits geometric conditions associated with true collision. Refinements that incorporate direct Earth-impact calculations, including gravitational focusing and atmospheric entry constraints, would enable more precise assessment of warning times and impact-specific observability. Such population would allow us to further study how the location of the eventual impact correlates with detectability. 

    \item \textit{Limited diversity in orbital parameter space:}
    Although our synthetic impactors inherit their orbital elements from the NEOMOD3-derived NEO distribution, additional effort is needed to ensure full coverage of rare but high-consequence dynamical classes—e.g., extremely low-inclination impactors, long-synodic-period Atiras, and retrograde impactors. Injecting targeted priors or importance sampling in these regions will allow a more complete mapping of survey performance across the full hazard space.
    \item \textit{Reliance on LSST \texttt{baseline\_v3.4} rather than the final survey strategy:}
    Our analysis uses the \texttt{baseline\_v3.4} cadence as a representative LSST survey realization. Although this cadence captures the general survey philosophy, it does not reflect the final adopted strategy, nor does it encompass the full range of scheduler updates, rolling cadence tests, or nightly optimization that the Rubin Observatory may ultimately implement. Future work should compare multiple OpSim strategy families, incorporate the final survey plan and evaluate the robustness of our conclusions across realistic strategy variations.

    \item \textit{Absence of a dedicated Argus Array moving object pipeline system:}
    Because the Argus Array does not yet have an operational moving-object processing pipeline comparable to those of Pan-STARRS \citep{Denneau_2013} and LSST \citep{Vere2017, Vereš_2017}, our results rely on an Argus-like detection threshold with LSST's linking analysis rather than a fully implemented end-to-end pipeline. A future priority will be to couple Argus pointing simulations with a dedicated moving-object processing pipeline to enable realistic cross-survey comparisons.

    \item \textit{Cross-survey linking and follow-up strategies:}
    Our analysis treats LSST and Argus as independent systems. In practice, the most consequential gains for planetary defense will arise from joint operations: cross-survey linking, rapid handoff of unlinked LSST candidates to Argus (or vice versa), and optimized follow-up using existing assets. Existing surveys already occupy complementary roles, with ATLAS providing continuous, all-sky, multi-nightly coverage for small imminent impactors \citep{Tonry2018}, Pan-STARRS combining moderate depth with single-night tracklets submission capability \citep{Chambers2016, Denneau_2013} and forthcoming infrared assets such as NEO Surveyor extending sensitivity to low–solar-elongation geometries inaccessible from the ground \citep{mainzer2023nearearthobjectsurveyormission}. Rubin Observatory In-Kind Program explicitly anticipates such integration of external survey and follow-up capabilities into LSST science operations.\footnote{See \url{https://www.lsst.org/scientists/in-kind-program}.} Simulations that include inter-survey cadence coordination, multi-system linking pipelines, and triggered follow-up observations represent a major avenue for future improvement.
    \item \textit{Toward operational end-to-end simulations:}
    Finally, a natural extension of this work is the development of full end-to-end simulations that propagate objects from synthetic generation to survey detection, linking, orbit determination, and impact probability assessment. Such pipelines will enable us to quantify not only detectability, but also the reliability and timeliness with which hazardous objects can be confirmed.
\end{itemize}

At both the population and individual-object level, these developments will enable quantitative assessment of full-network readiness for planetary defense.

\section{Conclusions}
\label{sec:conclusions}
We present a survey-realistic, population-scale assessment of the Vera C.\ Rubin Observatory’s ability to discover Earth-impacting asteroids prior to collision. Because existing LSST simulations have not explicitly focused on large-scale Earth-impacting populations, we construct a new synthetic impactor population by minimally perturbing \texttt{NEOMOD3}-derived NEOs and propagate them through the \texttt{Sorcha} survey simulator.

LSST’s performance is size-dependent. It discovers 79.7\% of impactors larger than 140 m, but completeness drops to 50.3\% for 50--140 m objects and 26.8\% for 20--50 m Tunguska-scale bodies, and 10.5\% for 10-20 m Chelyabinsk-scale objects, with impactors $<50$ m are typically discovered only weeks before impact, and 10.1\% of 50-140 m and 39.0\% $>140$ m impactors are discovered more than one year in advance.

A loss-mode analysis reveals that LSST’s incompleteness arises from photometric and linking constraints. Small objects are limited by photometric reach, while moderate and large impactors are missed primarily due to linking. Many such objects are observed above the limiting magnitude yet fail to obtain subsequent nights of observations required by linking process. Linking constraints account for 34\% of the total losses, making LSST fundamentally depth-rich but cadence-limited.

To evaluate complementary architectures, we conduct a conceptual simulation of an Argus-like, high-cadence all-sky survey. Although shallow, Argus provides dense temporal sampling that effectively eliminates linking failures for objects above its magnitude limit, recovering unlinked impactors by LSST. Conversely, LSST discovers the faint, distant, long-warning population that Argus cannot access. Together they provide a more complete coverage of imminent-impact trajectories.

These results demonstrate the challenges to provide long-lead discovery across the hazardous size spectrum by LSST alone. A coordinated, multi-survey architecture that combines LSST’s depth with high-cadence all-sky monitoring is essential for achieving robust planetary-defense readiness in the Rubin era.

\section*{Acknowledgments}

We thank all the people who contributed valuable discussions to this work. 

Q.C., M.B.F. and D.S. acknowledge support from the Duke University Trinity College of Arts and Sciences Department of Physics and from the Cosmology Group.

M.B.F. and D.S. acknowledges support from the Duke University Electrical and Computer Engineering Department.

D.S. is supported by Department of Energy grant DE-SC0010007, the David and Lucile Packard Foundation, the Templeton Foundation, and Sloan Foundation.

J.A.K. acknowledges the support from the University of Washington College of Arts and Sciences Department of Astronomy and thanks the LSST-DA Data Science Fellowship Program, which is funded by LSST-DA, the Brinson Foundation, and the Moore Foundation; his participation in the program has benefited this work. 

J.K. and I.C. acknowledge support from the DIRAC Institute in the Department of Astronomy at the University of Washington. The DIRAC Institute is supported through generous gifts from the Charles and Lisa Simonyi Fund for Arts and Sciences, and the Washington Research Foundation.


\appendix

\section{Detailed Sorcha configuration for component-wise assessment}
\label{appendix:sorcha_configuration}

To assess how individual survey-performance components shape impactor recover ability, we constructed a sequence of Sorcha configurations (in1–in4) that progressively enable successive stages of the Sorcha processing pipeline in order to isolate the dominant sources of discovery loss (configuration shown in Table~\ref{tab:sorcha_config_incre}). The first configuration (``in1") includes only single-exposure photometric modeling—trailed apparent magnitudes, trailing losses, vignetting (2 sigma SNR cuts and the camera footprint will be applied by default after including these filtering)—thereby probing fundamental per-visit detectability. In ``in2", we then introduce observation randomization to account for LSST’s pointing history and sky coverage. Next, we enable Sorcha’s source-detection efficiency (“fading”) function in ``in3", which probabilistically removes marginal detections near the limiting magnitude and isolates losses due to imperfect per-visit detection efficiency. Finally, we activate the linking filter to model the conversion of detections into discoveries (``in4").
\begin{table*}[ht!]
\centering
\caption{Incremental LSST Sorcha configuration used in diagnostic experiments. 
Each successive configuration adds one additional survey filter or noise model.}
\label{tab:sorcha_config_incre}
\begin{tabular}{l c c c c}
\hline\hline
\textbf{Filter / Model} & \textbf{in1} & \textbf{in2} & \textbf{in3} & \textbf{in4 (full)}\\
\hline
Trailing loss and bright limit& ON & ON & ON & ON \\
Vignetting model            & ON  & ON  & ON  & ON  \\
Randomization            & OFF & ON & ON & ON \\
Fading function           & OFF & OFF & ON & ON \\
Linking & OFF & OFF & OFF & ON \\
\hline
\end{tabular}
\end{table*}

\section{Early-Survey Impact-Time Bias Tests}
\label{appendix:early_bias}

To assess the impact of intrinsically truncated discovery windows during the earliest phase of survey operations, we repeat the full discovery-rate and warning-time analysis after excluding all impactors with impact dates prior to 2027 May 1. This restriction removes objects whose maximum achievable warning time is limited by the survey start epoch.

After excluding these early impacts, discovery efficiencies remain strongly size dependent, rising from 9.9\% for 10–20 m objects to 25.4\% for 20–50 m, 48.9\% for 50–140 m, and 80.9\% for impactors larger than 140 m. Comparing to the full impactor population, such removal only changes discovery rate at the few-percent level across all size bins (-5.4\% for 10-20 m, -5.1\% for 20-50 m, -2.8\% for 50-140 m, and +1.5\% for $>140$ m), indicating that early-survey impactors do not significantly bias overall discovery. Median warning times increase with size: no measurable change (+0.0 days) for 10-20 m objects, a +1.5 day increase for 20-50m impactors, a +15 day increase for 50-140 m bodies, and a +233 day increase for objects larger than 140 m. This pattern reflects the preferential removal of intrinsically time-limited early impacts-particularly among large objects rather than any change in the survey's intrinsic sensitivity. 

\begin{figure*}[ht!]
\centering
\includegraphics[width=0.7\textwidth]{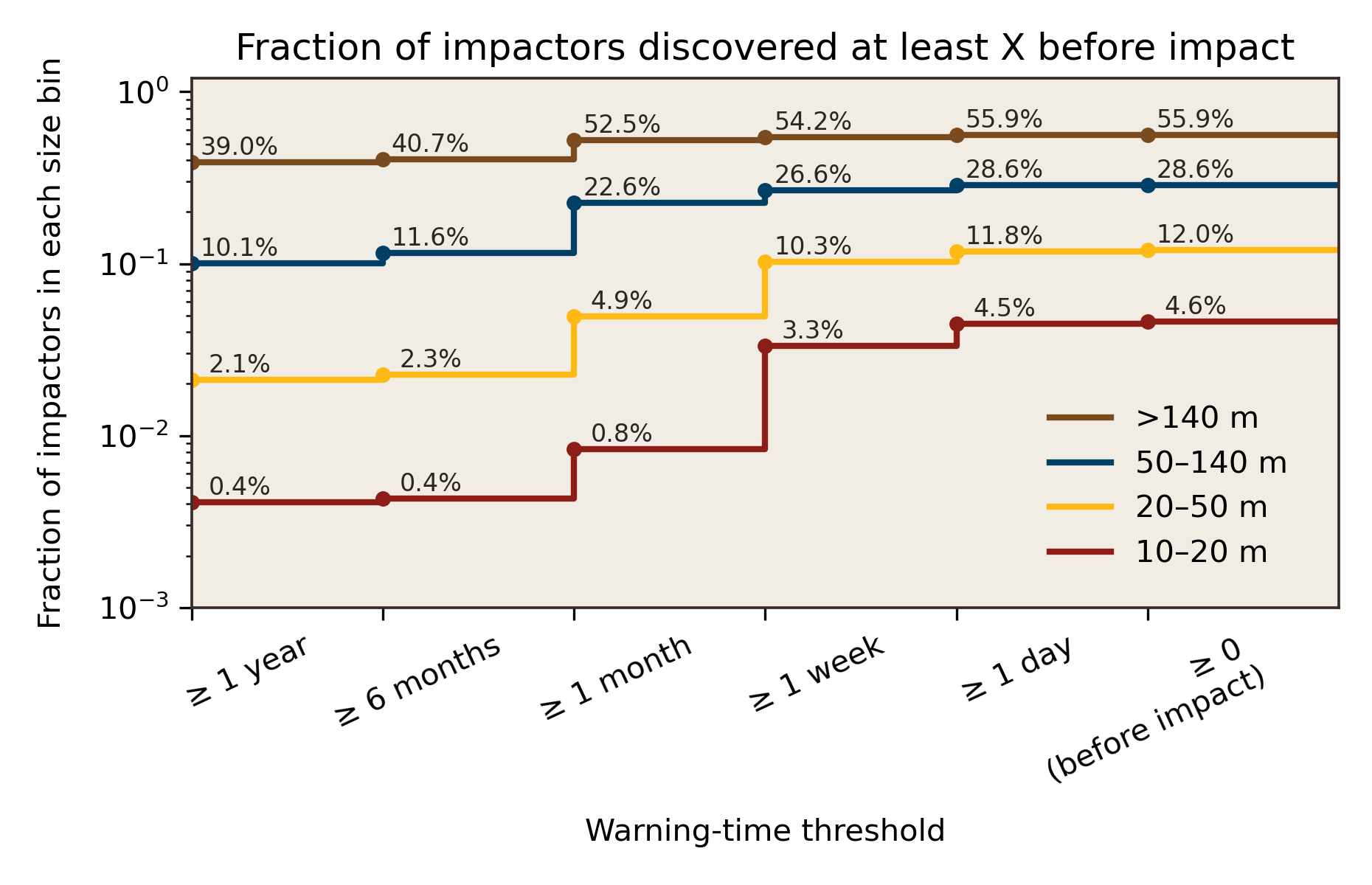}
\caption{
Cumulative fraction of impactors discovered at least $X$ days before impact for the subset of objects impacting on or after 2027 May~1, shown for four size bins (10--20~m, 20--50~m, 50--140~m, and $>140$~m).
\label{fig:early_bias_warn_cdf}}
\end{figure*}

For completeness, Figure~\ref{fig:early_bias_warn_cdf} presents the cumulative warning-time distributions for each size bin for the restricted sample.

\section{Discovery-Circumstance Diagnostics}

\begin{figure*}[!t]
\centering

\gridline{
\fig{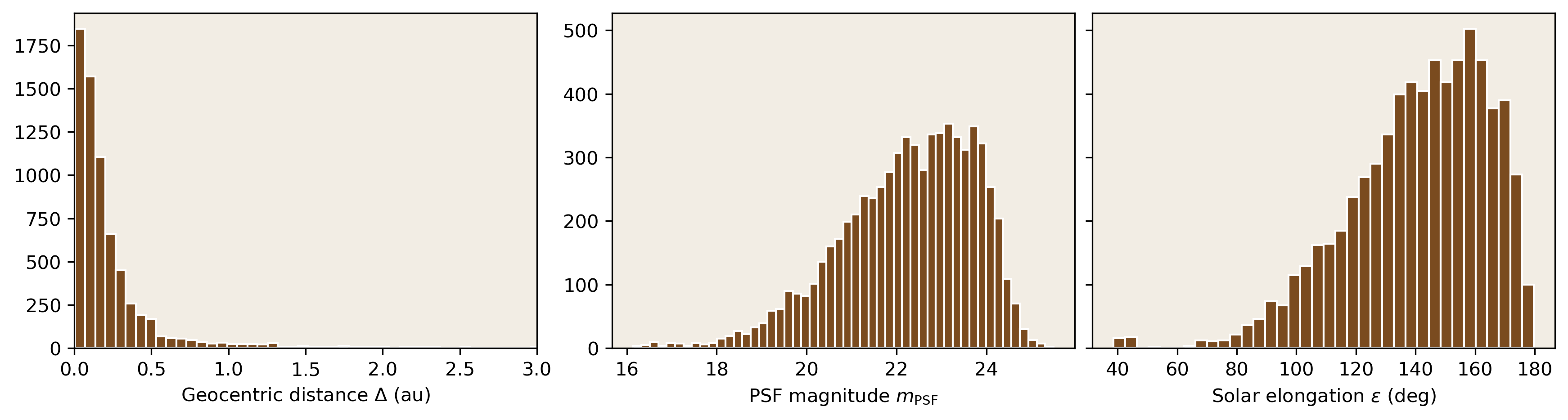}{\textwidth}{\textbf{(a) Geometric.}}
}

\vspace{0.5em}

\gridline{
\fig{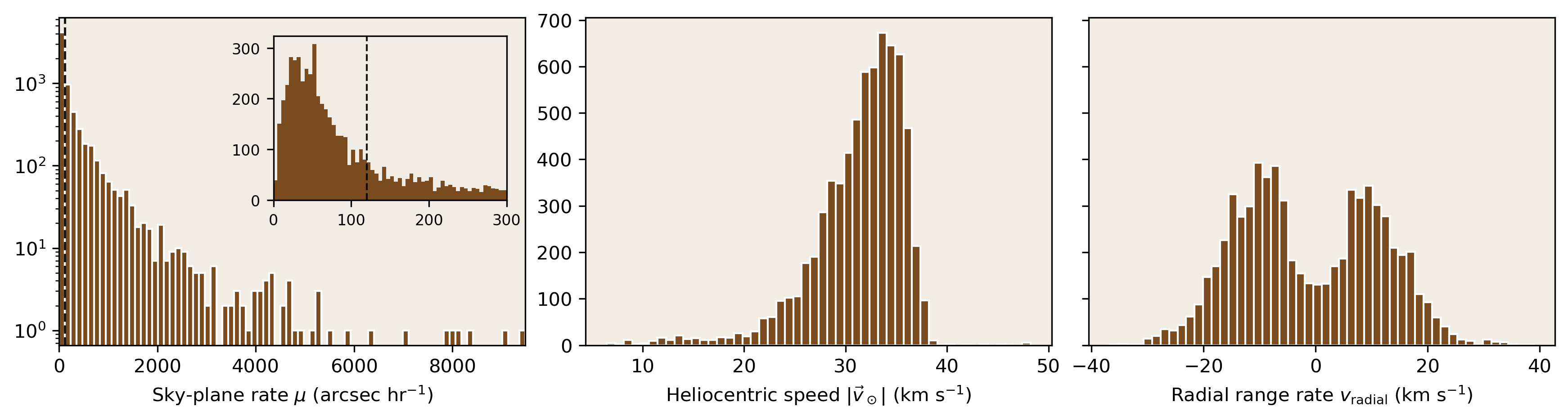}{\textwidth}{\textbf{(b) Kinematic.}}
}

\vspace{0.5em}

\gridline{
\fig{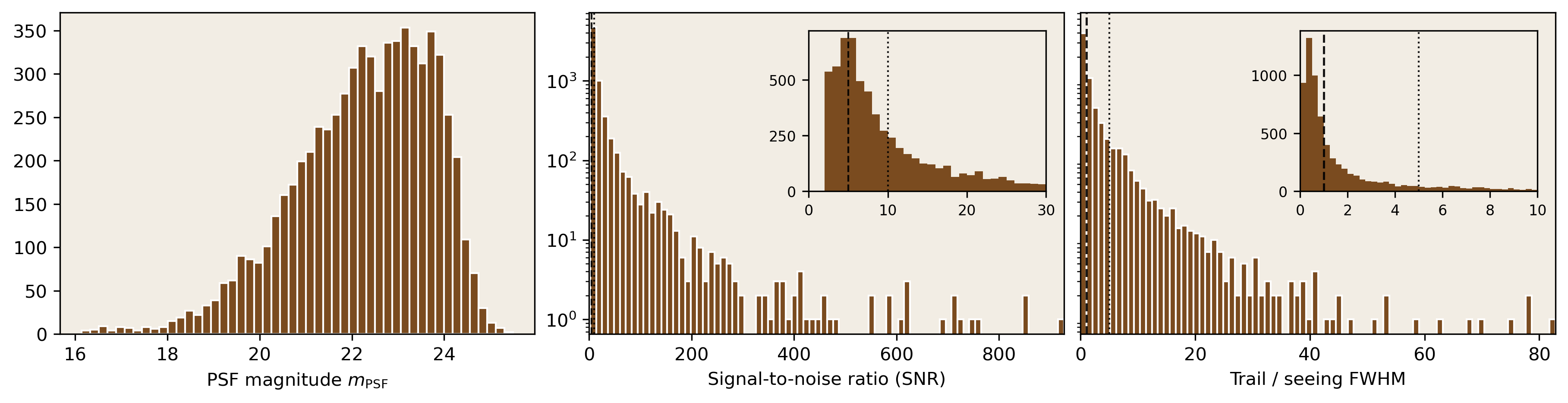}{\textwidth}{\textbf{(c) Photometric.}}
}

\caption{
Discovery diagnostics for discovered impactors at the moment of their first LSST detection.
\textbf{(a) Geometric:} distributions of geocentric distance $\Delta$, solar elongation $\epsilon$, and phase angle $\alpha$.
\textbf{(b) Kinematic:} sky-plane rate $\mu$ with zoomed-in panel, heliocentric speed $\lvert \vec{v}_{\odot} \rvert$, and radial range rate $v_{\rm radial}$.
\textbf{(c) Photometric:} $m_{\rm PSF}$, SNR with zoomed-in panel, and trail-to-seeing ratio with zoomed-in panel.
}
\label{fig:appendix_discovery_diagnostics}
\end{figure*}

Figure~\ref{fig:appendix_discovery_diagnostics} summarizes the distributions of (a) geometric, (b) kinematic, and (c) photometric properties at first detection. Impactors are typically detected at comparatively small geocentric distances $\Delta \lesssim 0.1$~AU. Smaller impactors are generally only discovered when they are already very close to Earth: those under 50 m are first detected at a median distance of $\sim$0.1 au (10th–90th percentile: 0.0–0.3 au). Mid-sized objects (50–140 m) appear at a median of $\sim$0.6 au (0.1–1.0 au), while the largest impactors ($>$140 m) are typically detected much farther out, with a median of $\sim$1.1 au and a broad 0.5–2.7 au range. Solar elongations are strongly concentrated at large values ($\epsilon \gtrsim 120^\circ$), indicating that pre-impact discoveries are made predominantly on the evening and morning sky near opposition rather than at low elongation.  The corresponding phase angles span $\alpha \sim 10^\circ$--$60^\circ$, implying a substantial loss of reflected-light brightness relative to full illumination.  Thus, LSST generally first detects impactors when they are already relatively close to Earth but only partially illuminated.

In terms of kinematic properties in panel (b), most first detections occur at modest sky-plane rates, with a strong concentration at $\mu \lesssim$ a few hundred arcsec~hr$^{-1}$, so that the resulting trails in a single LSST exposure are comparable to or smaller than the seeing disk.  The inferred transverse velocities relative to Earth are typically only a few km~s$^{-1}$, much smaller than the heliocentric speeds $|\vec{v}_\odot| \simeq 25$--$40$~km~s$^{-1}$. By comparing $v_\perp$ and $|\vec{v}_\odot|$ directly, the apparent motion on the sky is only a small fraction of the true orbital speed.  The line-of-sight range rates are correspondingly large and predominantly inbound, as expected when selecting only pre-impact discoveries.  Together, these trends indicate that LSST usually first observes impactors while they are moving largely along the line of sight rather than sweeping rapidly across the sky. 

Panel (c) characterizes the photometric state of impactors at first detection. The PSF magnitudes cluster around $m_{\rm PSF} \approx 22$--24, typically within $\sim$1--2~mag of the single-visit 5$\sigma$ depth, showing that most objects are detected slightly brighter than the nominal limiting magnitude. The corresponding SNR values are generally well above the detection threshold, indicating that many impactors are recovered with comfortable significance rather than at the very edge of sensitivity. Finally, the trail-to-seeing ratio is strongly peaked at values $<1$, confirming that trailing losses are modest for the vast majority of pre-impact discoveries.

\section{Discovery Rate Figures by stages of filtering}
\label{appendix: stage_filtering}

\begin{figure*}
\gridline{
  \fig{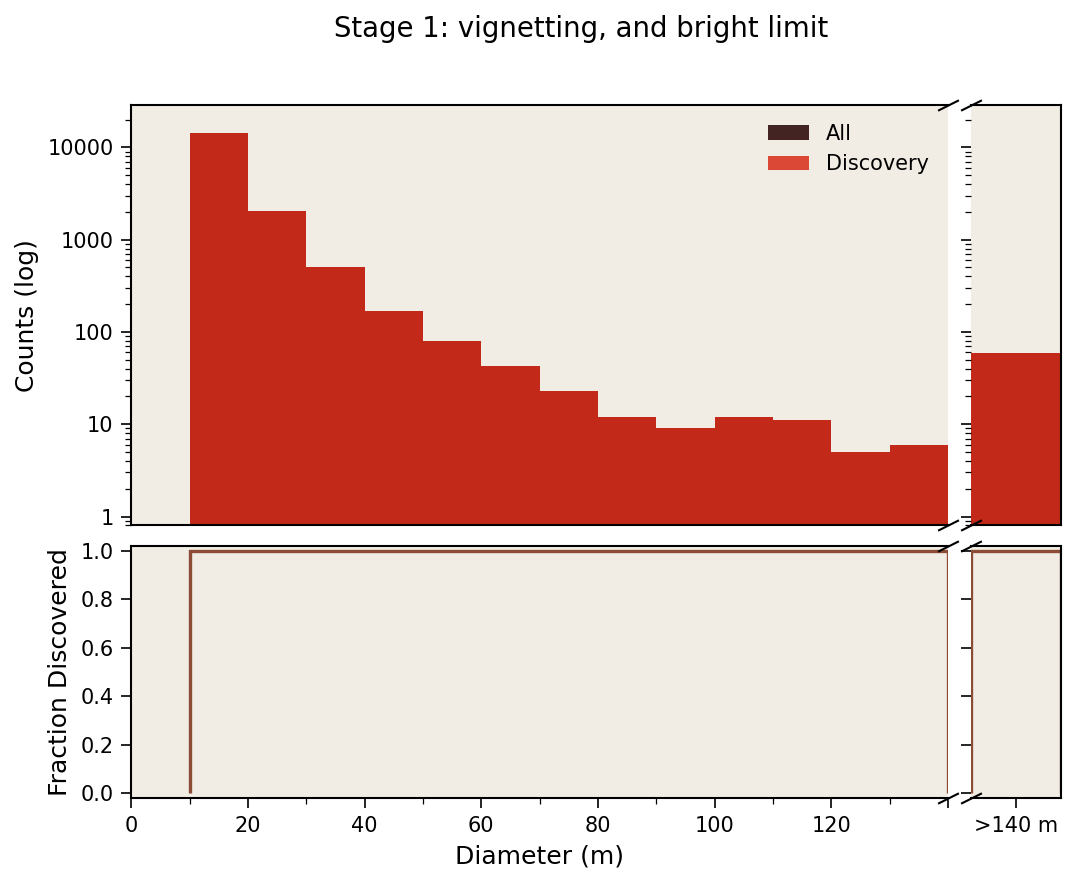}{0.47\textwidth}{(a)}
  \fig{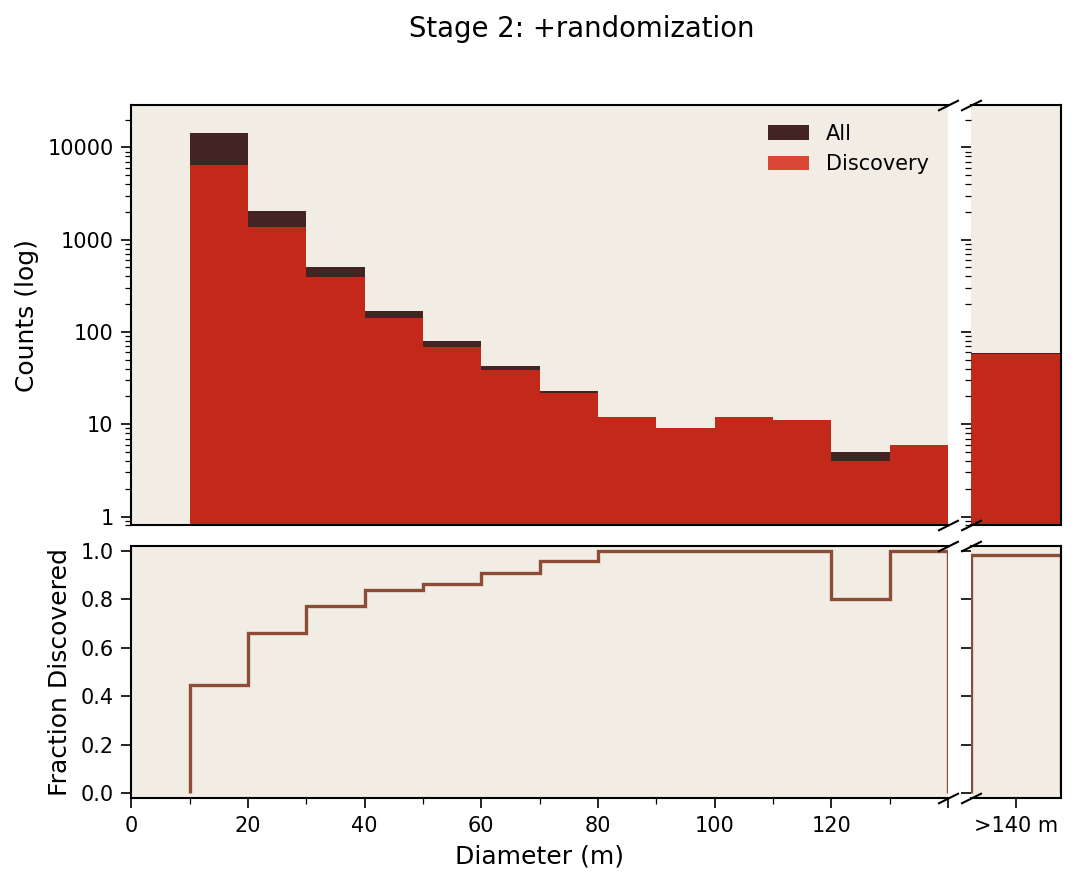}{0.47\textwidth}{(b)}
}
\gridline{
  \fig{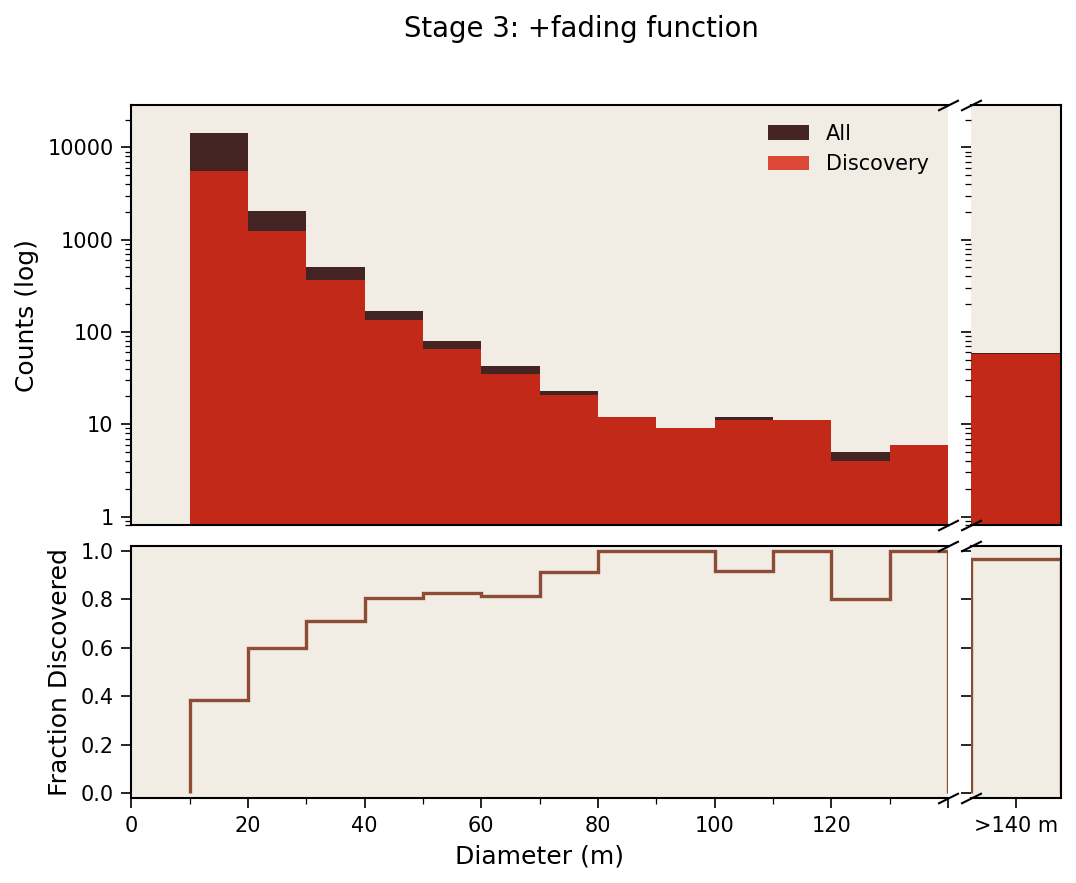}{0.47\textwidth}{(c)}
  \fig{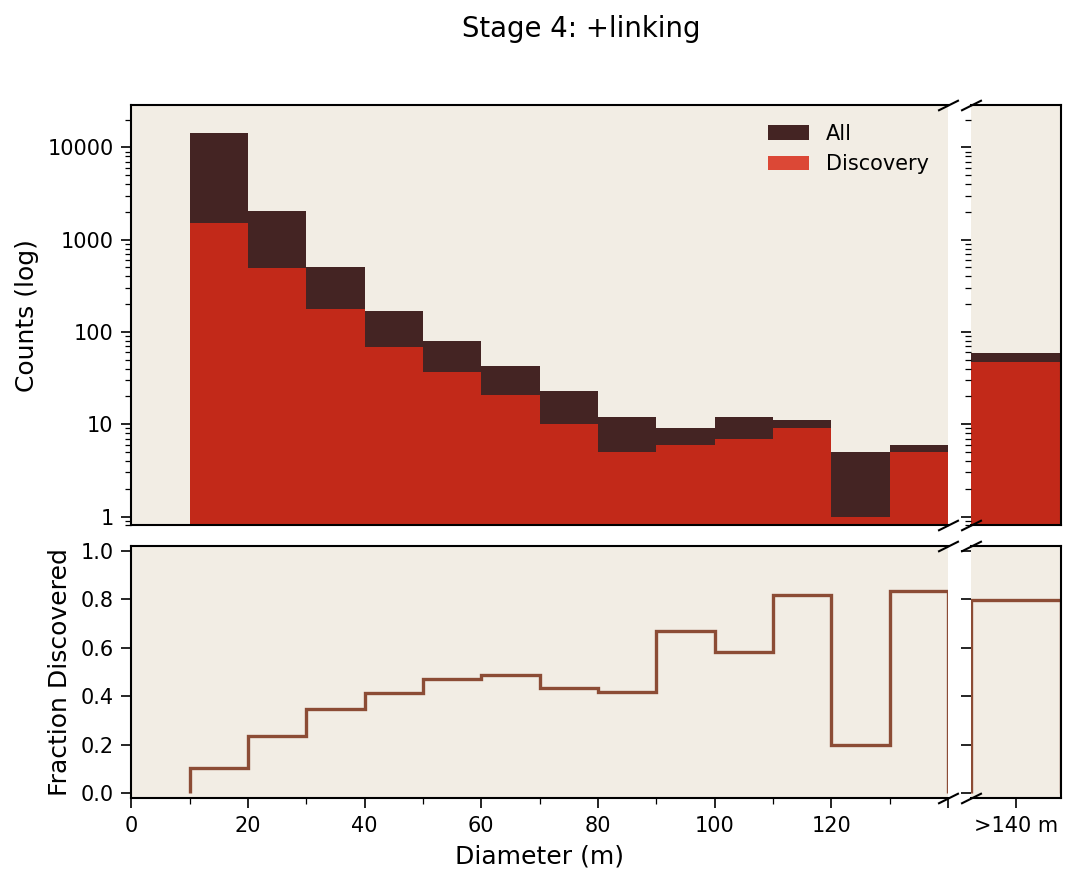}{0.47\textwidth}{(d)}
}
\caption{
Successive application of Sorcha’s survey-simulation filters and their impact on LSST impactor discovery. Each panel shows the differential size distribution (top) and the discovered-to-total ratio (bottom) as a function of impactor diameter. 
(a) Stage 1 includes only single-exposure photometric modeling (trailed magnitude, vignetting, bright-limit), isolating per-visit detectability. (b) Stage 2 adds photometric randomization, introducing realistic pointing history and sky coverage. (c) Stage 3 enables the source-detection efficiency (“fading”) function, which probabilistically removes marginal detections near the limiting magnitude. (d) Stage 4 activates the linking filter, requiring detections to form valid tracklets and tracks to be classified as discoveries. Together, the panels show how progressively enabling Sorcha filters transforms single-visit detectability into end-to-end survey discovery efficiency.
}
\label{fig:panel_lsst_stages}
\end{figure*}

\begin{figure*}
\gridline{
  \fig{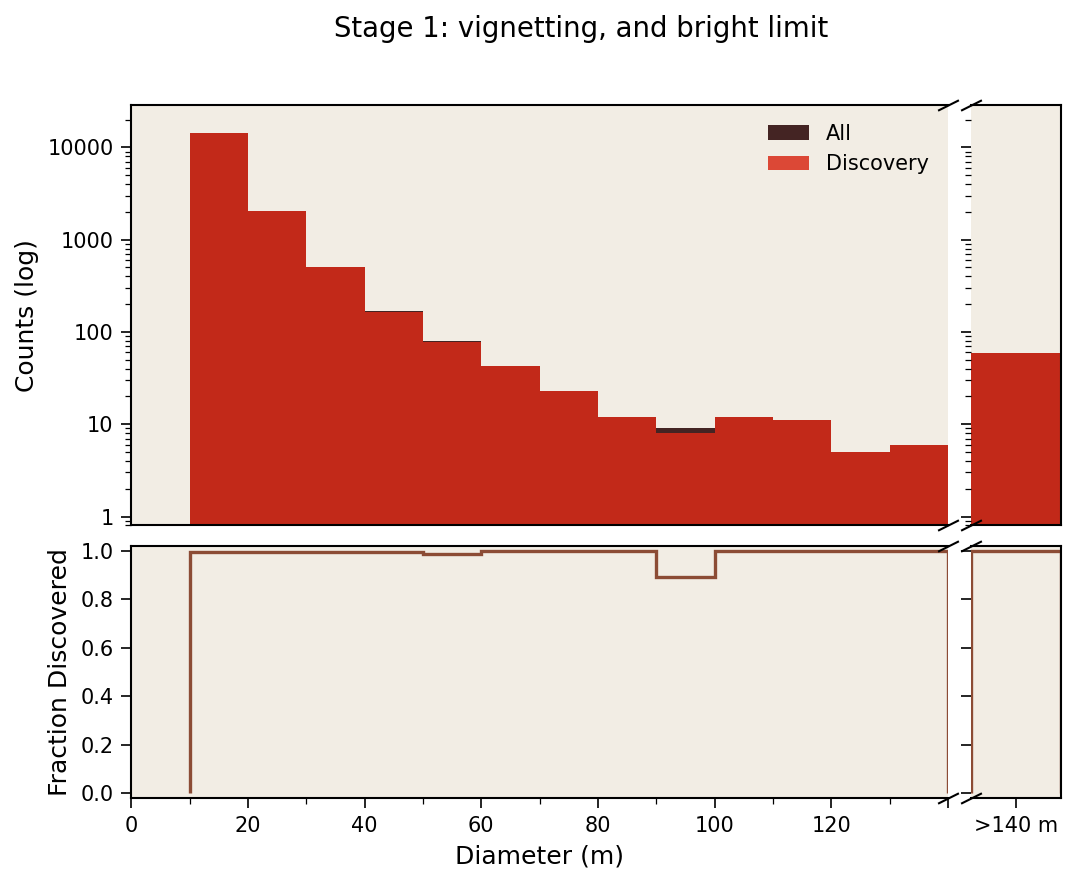}{0.47\textwidth}{(a)}
  \fig{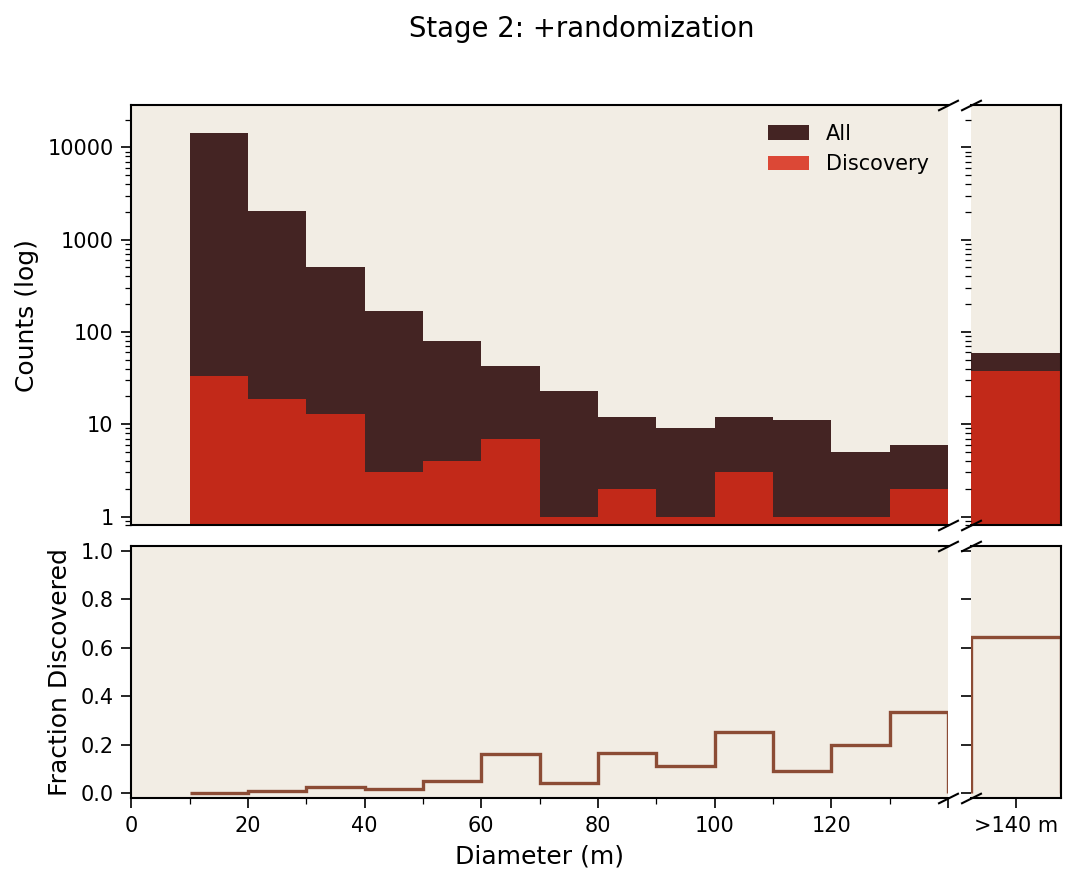}{0.47\textwidth}{(b)}
}
\gridline{
  \fig{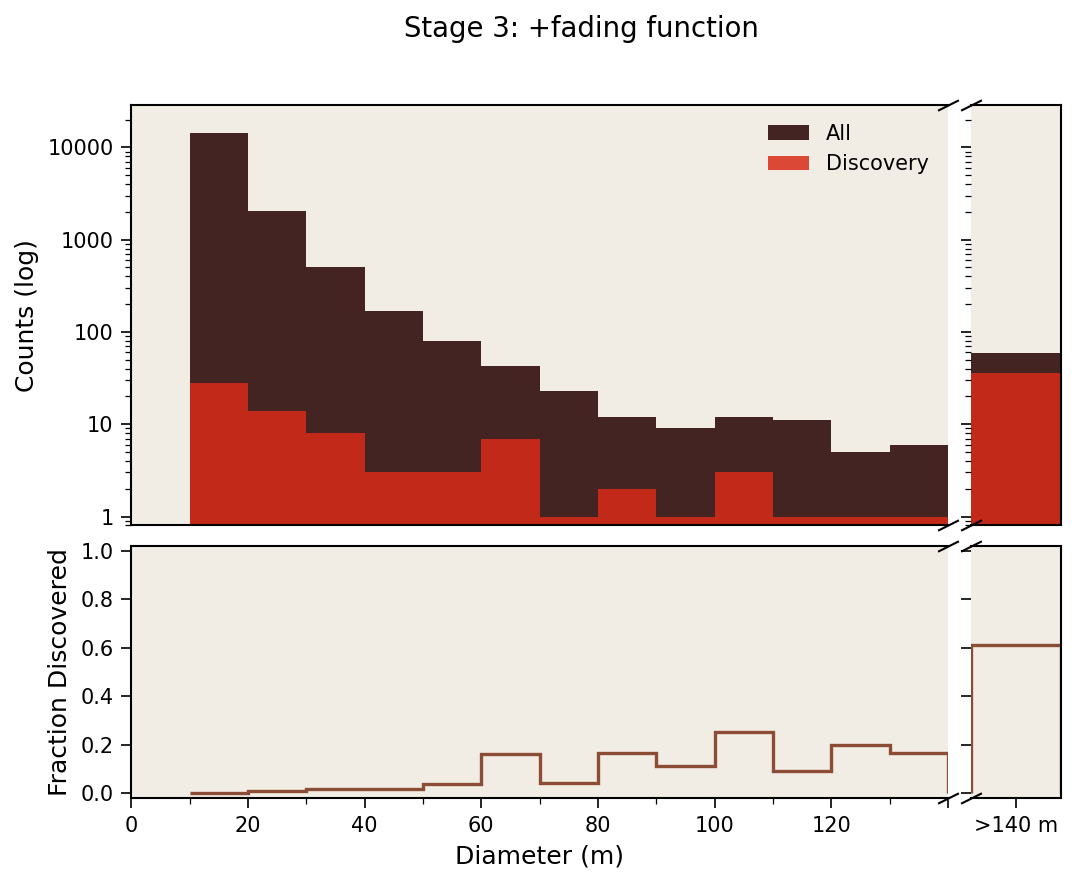}{0.47\textwidth}{(c)}
  \fig{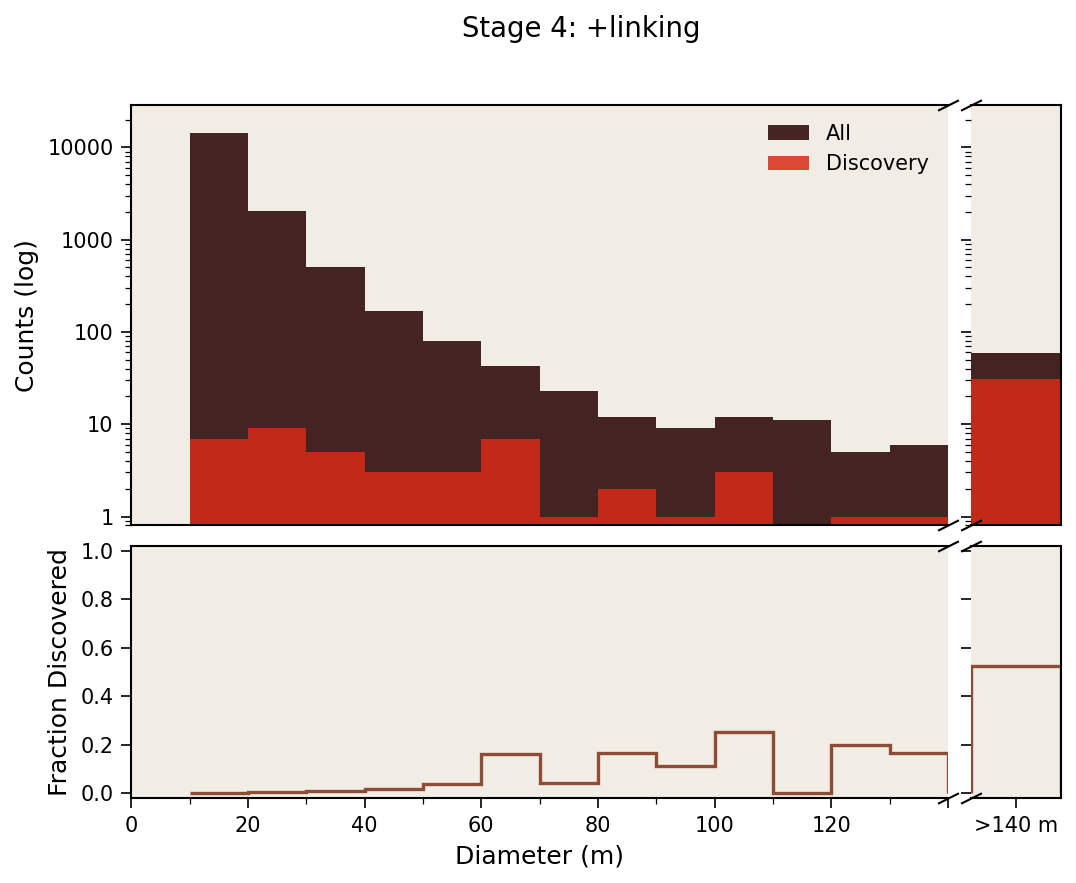}{0.47\textwidth}{(d)}
}
\caption{
Successive application of \texttt{Sorcha}'s survey-simulation filters and their impact on Argus impactor discovery, companion to Figures~\ref{fig:panel_lsst_stages}. Each panel shows the differential size distribution of all simulated impactors and the subset retained at that stage (top) and the corresponding retained fraction as a function of diameter (bottom). (a) Stage 1 applies single-exposure photometric modeling (vignetting, bright magnitude limit) (b) Stage 2 adds observation randomization, introducing realistic pointing and per-visit variability. (c) Stage 3 enables the source-detection efficiency (“fading”) function, probabilistically removing marginal detections near the limiting magnitude. (d) Stage 4 applies the linking filter, retaining only detections that form valid linkage. In contrast to the LSST sequence (Appendix~\ref{fig:panel_lsst_stages}), Argus losses are dominated by photometric limitations, with relatively few additional objects removed at the linking stage. Together, these panels show that Argus primarily recovers only the brightest, late-appearing impactors, while cadence play a secondary role.
}
\label{fig:argus_stages}
\end{figure*}

Figures~\ref{fig:panel_lsst_stages} and \ref{fig:argus_stages} illustrate the stage-by-stage evolution of impactor discovery for LSST and an Argus-like survey, based on the incremental configuration in \texttt{Sorcha} defined in Appendix~\ref{appendix:sorcha_configuration}. Together, these figures show how intrinsic detectability is transformed into end-to-end discovery efficiency, and how the relative importance of photometric and linking losses differs between LSST and Argus.

\bibliographystyle{aasjournal}
\bibliography{impactors}

\end{document}